\documentclass{article}

\input{tcilatex}
\begin{document}

\title{Gauge CPT as a Possible Alternative to
the Dark Matter Hypothesis}
\author{Kurt Koltko\\localcpt@yahoo.com}
\date{30 May 2013}
\maketitle

\begin{abstract}
A new force is proposed in order to explain galactic rotation curves. \ CPT
is chosen as the underlying symmetry of the new force because it is a
universal spacetime symmetry. \ Local CPT transformations are presented for
the Dirac field (matter) and the vierbein describing curved spacetime. \ A
nonvanishing variation of the Dirac action in curved spacetime is thus
derived. \ Because the metric spin connection of general relativity cannot
accommodate the variation induced by the local CPT symmetry, a new gauge
field is introduced. \ The transformation of the new field is derived which
implies the new field is massless. \ Experimental speculations based on the
zero mass of the new gauge field are presented. \ It is shown that one type
of Yang-Mills Lagrangian density is not invariant under local CPT
transformations.
\end{abstract}

\section{INTRODUCTION}
We gauge the CPT transformation in order to unveil a new spacetime dynamical
degree of freedom, i.e. a new force, with the hope this could shed some
light on current problems involving gravity. \ In particular, the galactic
"dark matter" problem will be addressed in this paper.

The basic idea is simple - instead of having unknown matter source a
required gravitational field, why not consider known matter as a source for
a new force? \ All of the currently known forces can be derived by gauging
certain continuous global symmetries, hence it would seem interesting to
gauge the CPT symmetry - even though it is a discrete global symmetry - just
to see what happens. \ We note that the CPT symmetry has been experimentally
verified and requires no more dimensions than the four which we know exist.
\ In other words, a natural basis exists for the notion of gauging CPT.

Specifically, it is the mass independent acceleration appearing in the
galactic rotation curves which suggests gauging CPT. \ First, CPT is a
universal symmetry as are the global proper orthochronous Lorentz
transformations which - when gauged - lead to the spin connection
formulation of general relativity [1].
\ Second, PT is also a proper Lorentz
transformation which suggests that it should be included in gauging the full
Lorentz group. \ These two characteristics shared with the spin connection
formulation of general relativity suggest that if a new force is uncovered
by gauging CPT, then it would obey the principle of equivalence - a mass
independent acceleration will occur.

CPT is particularly intriguing because it is the offspring of the
phenomenologically successful union - quantum field theory - of the global
theory of special relativity with quantum mechanics. \ By gauging CPT along
with special relativity, perhaps we can elevate the status of CPT to that of
a "bridge" needed for the unification of general relativity with quantum
theory by unveiling additional spacetime dynamical degrees of freedom.

\section{THE TRANSFORMATIONS}
At first glance it may not appear possible to gauge CPT because there are no
important continuously varying parameters involved with the CPT
transformation. \ A $U(1)$ phase could be included in the CPT transformation
of the Dirac spinor $\psi $ ; however, we ignore this because it can be
absorbed into the $U(1)$ gauge transformation associated with the
electroweak interactions. \ Locality would also seem to be a problem. \
Except for an infinitesimal neighborhood around the origin of a Minkowski
manifold, P and T are not local transformations.

We examine the CPT transformation at the origin of an inertial reference
frame in order to overcome the above obstacles. \ The effect of the global
CPT transformation at the origin of a Minkowski spacetime coordinate system
is to "flip" the coordinate axes and transform a Dirac wavefunction from $%
\psi $ to $i\gamma ^{5}\psi $ (we are using Bjorken-Drell conventions).
 \ If a nontrivial \textit{spacetime } analog
of the charge conjugation operation exists, then we would have to include
its effect. \ We assume no such spacetime operation exists:  $C=I$, where $I$
is just the identity, \textit{for} \textit{spacetime} \textit{only}.  
In other words, we assume there is no such thing as an "antispacetime"
 distinct from spacetime. \ An
attempt to find a nontrivial spacetime $C$ operation is contained in [2]. \
To picture what is going on, we introduce a vierbein field $e_{a}^{\;\mu }$,
where $\mu $ represents the manifold coordinates and $a$ represents the
local inertial frame coordinates. \ We define a local CPT transformation as
the application of these "origin transformations" to vierbein and
wavefunctions at arbitrarily chosen points in a pseudo-Riemannian spacetime
manifold.

The choice of \textit{where} we want to perform a local CPT transformation
will, in part, play the role of the arbitrarily chosen continuous parameters
appearing in gauge theories. \ In order to make this concept precise, we
introduce a real scalar function, $f\in C^{1}$, defined over the entire
manifold to be used as the argument of step functions $\Theta $. \ In the
arbitrary regions where we choose to perform the local CPT transformations,
we set $f>0$ so that $\Theta \left[ f\right] =1$. \ In the arbitrary regions
where we choose not to perform local CPT transformations, we set $f<0$ so
that $\Theta \left[ -f\right] =1$. \ The boundaries between regions where
local CPT is carried out and where it is not are given by $f=0$ with the
convention that $\Theta \left[ f\right] =0$ if $f\leq 0$.

We emphasize that $f$ \ is \textit{not }a physical field. \ The step
functions, $\Theta $, are parameters which define when the local CPT
transformations are carried out (or not). \ The $\Theta $ are just like the $%
\lambda $ used in the gauging of $U\left( 1\right) $ except that there are
only two choices regarding the discrete CPT symmetry instead of the
continuum of choices for $\lambda $ to be used in the $U\left( 1\right) $
symmetry operation $e^{i\lambda }$. \ To make the $U\left( 1\right) $
operation local, one makes $\lambda $ an \textit{arbitrary} function of
spacetime subject only to the condition that $\lambda $ is differentiable. \
Similarly, the function $f$ \ is introduced in order to make the arbitrary
choice of carrying out local CPT $\left( f>0\right) $ or not $\left(
f<0\right) $ at the points of interest. \ The function $f$ \ plays the same
role as replacing a constant $\lambda $ by $\lambda \left( x\right) $. \ The
only restriction placed on$\ f$ is that it be differentiable so that $%
\partial _{\mu }\Theta \left[ \pm f\right] =\pm \delta \left[ f\right]
\partial _{\mu }f$ \ makes sense ($\delta $ being the Dirac delta
functional).

Because we are utilizing the proper spacetime transformation PT, it would be
prudent to see if the metric spin connection $\omega _{\mu ab}$ alone could
accommodate local CPT transformations. \ To this end, we also include local
proper Lorentz rotations wherever $f>0$. \ The Lorentz rotations, $\Lambda $%
, are denoted by $\Lambda _{a}^{\;b}$ and $\Lambda _{\psi }$ for the
vierbein and Dirac wavefunction respectively. \ In effect, we are gauging
the $CPT\Lambda $ transformation of the Dirac field to induce the gauging of
the full group of proper spacetime Lorentz transformations.

Putting all of the above together, we have the following local $CPT\Lambda $
transformations:%
\begin{eqnarray}
e_{a}^{\;\mu } &\rightarrow &\Theta \left[ -f\right] e_{a}^{\;\mu }-\Theta %
\left[ f\right] e_{b}^{\;\mu }\Lambda _{a}^{\;b}, \\
\psi &\rightarrow &\Theta \left[ -f\right] \psi +\Theta \left[ f\right]
i\gamma ^{5}\Lambda _{\psi }\psi , \\
\overline{\psi } &\rightarrow &\Theta \left[ -f\right] \overline{\psi }%
+\Theta \left[ f\right] i\overline{\psi }\gamma ^{5}\Lambda _{\overline{\psi 
}}, \\
\omega _{\mu ab} &\rightarrow &\Theta \left[ -f\right] \omega _{\mu
ab}+\Theta \left[ f\right] \widetilde{\omega }_{\mu ab}+\delta \left[ f%
\right] \Theta \left[ -f\right] \varsigma _{\mu ab}+\delta \left[ f\right]
\Theta \left[ f\right] \widetilde{\varsigma }_{\mu ab},
\end{eqnarray}%
where $$\omega _{\mu ab}=\frac{1}{2}e_{a}^{\;\nu }\left( \partial _{\mu
}e_{b\nu }-\partial _{\nu }e_{b\mu }\right) -\frac{1}{2}e_{b}^{\;\nu }\left(
\partial _{\mu }e_{a\nu }-\partial _{\nu }e_{a\mu }\right) -\frac{1}{2}%
e_{a}^{\;\rho }e_{b}^{\;\sigma }\left( \partial _{\rho }e_{r\sigma
}-\partial _{\sigma }e_{r\rho }\right) e_{\;\mu }^{r}$$ and
 $\widetilde{\omega }%
_{\mu ab}$ is the transformation of $\omega _{\mu ab}$ under $CPT\Lambda $
and $\varsigma _{\mu ab}$, $\widetilde{\varsigma }_{\mu ab}$ are boundary
terms arising from the differentiation of the vierbein transformations in
the metric spin connection. \ The explicit expressions for $\widetilde{%
\omega }_{\mu ab}$, $\varsigma _{\mu ab}$, and $\widetilde{\varsigma }_{\mu
ab}$ are in appendix B.  The coordinate axes "flip" is given by the
 $-\Theta \left[ f \right] $ in eq. (1). \ The volume element transforms as $%
ed^{4}x\rightarrow \left( \Theta \left[ f\right] +\Theta \left[ -f\right]
\right) ed^{4}x$, where $e=\det \left( e_{\;\mu }^{a}\right) $. \ Clearly,
these transformations are well defined in curved spacetime.

If one feels uncomfortable with the defining transformation equations 1-4
because they appear unphysical (due to the discontinuities), then one need
only look at the usual gauge theories to see that those defining
transformation equations are also unphysical - hence the need to introduce
the compensating (i.e., gauge) fields. \ We return to the example of $%
U\left( 1\right) $ to illustrate how this occurs. \ We start with $\psi
\rightarrow e^{i\lambda \left( x\right) }\psi $ and examine $\left\langle
\psi \left\vert p_{x}\right\vert \psi \right\rangle $, where $p_{x}$ is the
momentum operator in the $x$ dimension. \ Under the $U\left( 1\right) $
transformation we see that:%
\[
\left\langle \psi \left\vert p_{x}\right\vert \psi \right\rangle \rightarrow
\left\langle e^{i\lambda \left( x\right) }\psi \left\vert -i\hbar \frac{%
\partial }{\partial x}\right\vert e^{i\lambda \left( x\right) }\psi
\right\rangle =-i\hbar \left\langle \psi \left\vert i\frac{\partial \lambda 
}{\partial x}\right\vert \psi \right\rangle +\left\langle \psi \left\vert
p_{x}\right\vert \psi \right\rangle . 
\]%
Now, if one considers the specific case of $\psi $ representing a free
particle without any forces present, then we see that the term $-i\hbar
\left\langle \psi \left\vert i\frac{\partial \lambda }{\partial x}%
\right\vert \psi \right\rangle $ introduces variations in the momentum. \ In
other words, the free particle undergoes arbitrary changes in its motion
without any forces present - clearly an unphysical situation.

The presence of discontinuities in the defining transformations are just a
reflection of the fact that the $CPT$ symmetry is a discrete symmetry rather
than a continuous symmetry. \ It is important to accept that the $CPT$
transformation is discrete, handle accordingly, and be aware of possible
novel terms arising in the free-field Lagrangian due to the discrete nature
of the transformations.

To make sense of such expressions, we demand that discontinuities containing 
$\delta \left[ f\right] $ disappear from the Lagrangian (i.e., the action 
\textit{integral}). \ The resulting field equations - and ensuing physical
predictions - will therefore be free of discontinuities. \ This requirement
is exactly analogous to requiring gauge invariance of expressions appearing
in the Lagrangians of other gauge theories\footnote{The free-field term,
 $Tr\left(
F^{\mu \nu }F_{\mu \nu }\right) $, in these theories transforms gauge
covariantly as $Tr\left( \Omega F^{\mu \nu }F_{\mu \nu }\Omega ^{-1}\right) $%
. \ By the cyclic property of traces, however, this is equal to $Tr\left(
F^{\mu \nu }F_{\mu \nu }\right) $.}. \ Indeed, returning to $\omega _{\mu
ab} $, it will be shown that the free-field Lagrangian (the scalar curvature 
$R$) constructed from $\omega _{\mu ab}$ is invariant under local $%
CPT\Lambda $ transformations up to removable singularities occurring where $%
f=0$. \ These particular singularities have no effect on the action and so
can be ignored.

\section{VARIATION OF THE DIRAC ACTION}
We begin with the Hermitian form of the Dirac Lagrangian density which gives
us the unvaried action:%
\begin{equation}
S=\int \left\{ \frac{i}{2}\left[ \overline{\psi }\gamma ^{a}e_{a}^{\;\mu
}\partial _{\mu }\psi -e_{a}^{\;\mu }\partial _{\mu }\overline{\psi }\gamma
^{a}\psi \right] -m\overline{\psi }\psi \right\} ed^{4}x,
\end{equation}%
where $m$ is the mass of the Dirac particle, $\gamma ^{a}$ are the Dirac
gamma matrices, and natural units are used. \ We apply transformations
(1)-(3) to eq. (5) to obtain the transformed action $S^{\prime }$. \ We note
that $\partial _{\mu }\left( \Theta \left[ \pm f\right] h\right) =\left( \pm
\delta \left[ f\right] \partial _{\mu }f\right) h+\Theta \left[ \pm f\right]
\left( \partial _{\mu }h\right) $, where $h$ is an "ordinary" function. \
Also, terms with coefficients $\Theta \left[ -f\right] \Theta \left[ f\right]
$ and $\Theta \left[ -f\right] \Theta \left[ f\right] \delta \left[ f\right] 
$ integrate to $0$ (independent of the convention for $\Theta \left[ 0\right]
$) and are dropped.  Appendix A discusses how to handle various terms 
containing products of delta functionals and step functions \ We obtain:%
\begin{eqnarray*}
S^{\prime } &=&\int \Theta \left[ -f\right] \left\{ \frac{i}{2}\left[ 
\overline{\psi }\gamma ^{a}e_{a}^{\;\mu }\partial _{\mu }\psi -e_{a}^{\;\mu
}\partial _{\mu }\overline{\psi }\gamma ^{a}\psi \right] -m\overline{\psi }%
\psi \right\} ed^{4}x \\
&&+\int \Theta \left[ f\right] \left\{ \frac{i}{2}\left( \partial _{\mu }%
\overline{\psi }\Lambda _{\overline{\psi }}+\overline{\psi }\partial _{\mu
}\Lambda _{\overline{\psi }}\right) e_{b}^{\;\mu }\Lambda _{a}^{\;b}\gamma
^{a}\Lambda _{\psi }\psi \right\} ed^{4}x \\
&&-\int \Theta \left[ f\right] \left\{ \frac{i}{2}\overline{\psi }\Lambda _{%
\overline{\psi }}\gamma ^{a}e_{b}^{\;\mu }\Lambda _{a}^{\;b}\left( \partial
_{\mu }\Lambda _{\psi }\psi +\Lambda _{\psi }\partial _{\mu }\psi \right)
\right\} ed^{4}x \\
&&+\int \Theta \left[ f\right] m\overline{\psi }\psi ed^{4}x \\
&&+\frac{i}{2}\int \delta \left[ f\right] \partial _{\mu }f\left\{ \Theta %
\left[ -f\right] ie_{a}^{\;\mu }\overline{\psi }\left( \gamma ^{a}\Lambda
_{\psi }+\Lambda _{\overline{\psi }}\gamma ^{a}\right) \gamma ^{5}\psi
\right\} ed^{4}x \\
&&-\frac{i}{2}\int \delta \left[ f\right] \partial _{\mu }f\left\{ \Theta %
\left[ f\right] ie_{b}^{\;\mu }\Lambda _{a}^{\;b}\overline{\psi }\left(
\Lambda _{\overline{\psi }}\gamma ^{a}+\gamma ^{a}\Lambda _{\psi }\right)
\gamma ^{5}\psi \right\} ed^{4}x.
\end{eqnarray*}

A couple of remarks are in order before calculating the variation of the
action $\delta S$. \ First, by setting $f>0$ everywhere, we see that the
local $CPT\Lambda $ transformations give the same form of the transformed
action $S^{\prime }$ as a global $CPT\Lambda $ transformation acting on the
action in Minkowski spacetime. \ Hence, we have a well defined transition
from local $CPT\Lambda $ to global $CPT\Lambda $ valid in curved spacetime.
\ Second, $S\rightarrow -S$ under global CPT in Minkowski spacetime. \
Therefore, the equations of motion for $\psi $ and $\overline{\psi }$ are
invariant under global CPT but the action is not. \ So, we must be careful
in defining $\delta S$ as $\delta S=S^{\prime }-S$. \ To obtain a vanishing $%
\delta S$ under global CPT transformations we would have to multiply $%
S^{\prime }$ by a factor of $-1$ before subtracting $S$. \ Therefore, we
multiply each volume integral occurring where $f>0$ in $S^{\prime }$ (i.e.
the $\Theta \left[ f\right] $ terms) by an additional factor of $-1$. \ We
do \textit{not} multiply the surface integrals containing $\Theta \left[ f%
\right] $ in $S^{\prime }$ (i.e. the $\Theta \left[ f\right] \delta \left[ f%
\right] $ terms) by $-1$ simply because there are no corresponding surface
integrals in $S$.

If one feels uncomfortable with the inclusion of the extra $-1$, then one
could leave it out and realize that $\delta S=S_{CPT}^{\prime }-S=-2S$ is
the signature of the global $CPT$ symmetry ($S_{CPT}^{\prime }$ being the
action obtained under global $CPT$). \ Any changes caused by the
introduction of the local $CPT\Lambda $ symmetry would result in $\delta
S=-2S+\delta S_{D}$ where the expression for $\delta S_{D}$ would be exactly
the same as obtained below. \ One then obtains:%
\begin{equation}
\delta S=\frac{i}{2}\int \Theta \left[ f\right] \left\{ \overline{\psi }%
\left( e_{b}^{\;\mu }\Lambda _{a}^{\;b}\Lambda _{\overline{\psi }}\gamma
^{a}\Lambda _{\psi }-e_{a}^{\;\mu }\gamma ^{a}\right) \partial _{\mu }\psi
\right\} ed^{4}x
\end{equation}
\begin{displaymath}
-\frac{i}{2}\int \Theta \left[ f\right] \left\{ \partial _{\mu }\overline{%
\psi }\left( e_{b}^{\;\mu }\Lambda _{a}^{\;b}\Lambda _{\overline{\psi }%
}\gamma ^{a}\Lambda _{\psi }-e_{a}^{\;\mu }\gamma ^{a}\right) \psi \right\}
ed^{4}x
\end{displaymath}
\begin{eqnarray*}
&&+\frac{i}{2}\int \Theta \left[ f\right] e_{b}^{\;\mu }\Lambda _{a}^{\;b}%
\overline{\psi }\left( \Lambda _{\overline{\psi }}\gamma ^{a}\partial _{\mu
}\Lambda _{\psi }-\partial _{\mu }\Lambda _{\overline{\psi }}\gamma
^{a}\Lambda _{\psi }\right) \psi ed^{4}x \\
&&-\frac{1}{2}\int \delta \left[ f\right] \partial _{\mu }f\Theta \left[ -f%
\right] e_{a}^{\;\mu }\overline{\psi }\left( \gamma ^{a}\Lambda _{\psi
}+\Lambda _{\overline{\psi }}\gamma ^{a}\right) \gamma ^{5}\psi ed^{4}x \\
&&+\frac{1}{2}\int \delta \left[ f\right] \partial _{\mu }f\Theta \left[ f%
\right] e_{b}^{\;\mu }\Lambda _{a}^{\;b}\overline{\psi }\left( \gamma
^{a}\Lambda _{\psi }+\Lambda _{\overline{\psi }}\gamma ^{a}\right) \gamma
^{5}\psi ed^{4}x.
\end{eqnarray*}

We now examine the volume integrals occurring in eq. (6). \ From
the identity $\Lambda _{a}^{\;b}\Lambda _{\overline{\psi }}\gamma
^{a}\Lambda _{\psi }=\gamma ^{b}$, we see that the first two volume
integrals vanish. \ The remaining integral containing $\partial _{\mu
}\Lambda _{\overline{\psi }}$ and $\partial _{\mu }\Lambda _{\psi }$ will
vanish upon the introduction of the metric spin connection term $\omega
_{\mu ab}\sigma ^{ab}$ as part of the covariant derivative acting on $%
\overline{\psi }$ and $\psi $, \ $\partial _{\mu }\psi \rightarrow \partial
_{\mu }\psi +\frac{1}{2}\omega _{\mu ab}\sigma ^{ab}\psi $ and $\partial
_{\mu }\overline{\psi }\rightarrow \partial _{\mu }\overline{\psi }-\frac{1}{%
2}\omega _{\mu ab}\overline{\psi }\sigma ^{ab}$, where $\sigma ^{ab}=\frac{1%
}{4}\left[ \gamma ^{a},\gamma ^{b}\right] $ (we are not using Bjorken-Drell
conventions for $\sigma ^{ab} $). \ Introducing the metric spin
connection into the covariant derivative results in an interaction term, $%
S_{\omega }=\frac{i}{4}\int e_{a}^{\;\mu }\omega _{\mu bc}\overline{\psi }%
\left\{ \gamma ^{a},\sigma ^{bc}\right\} \psi ed^{4}x$, which needs to be
added to eq. (5). \ The effect on the action due to the variation of the
metric spin connection under gauge $CPT\Lambda $ will be denoted by $\delta
S_{\omega }$.

The boundary integrals in eq. (6) can be simplified by again using the above
identity. \ Including the metric spin connection $\omega _{\mu ab}$, one
then obtains $\delta S=\delta S_{D}+\delta S_{\omega }$, where%
\begin{equation}
\delta S_{D}=-\frac{1}{4}\int \delta \left[ f\right] \partial _{\mu
}f\left\{ e_{a}^{\;\mu }\overline{\psi }\gamma ^{5}\left( \left[ \gamma
^{a},\Lambda _{\overline{\psi }}\right] +\left[ \Lambda _{\psi },\gamma ^{a}%
\right] \right) \psi \right\} ed^{4}x,
\end{equation}%
and%
\begin{eqnarray}
\delta S_{\omega } & = & \frac{i}{16}\int \delta \left[ f\right] \partial _{\mu }f
\overline{\psi }\left\{ \gamma ^{a},\sigma ^{bc}\right\} \eta
_{ad}e_{b}^{\;\mu }\left( \Lambda _{c}^{\;d}+\Lambda _{\;c}^{d}\right) \psi
ed^{4}x \\
&&-\frac{i}{16}\int \delta \left[ f\right] \partial _{\mu }f\overline{\psi }%
\left\{ \gamma ^{a},\sigma ^{bc}\right\} \eta _{bd}e_{c}^{\;\mu }\left(
\Lambda _{\;a}^{d}+\Lambda _{a}^{\;d}\right) \psi ed^{4}x \nonumber \\
&&+\frac{i}{16}\int \delta \left[ f\right] \partial _{\mu }f\overline{\psi }%
\left\{ \gamma ^{a},\sigma ^{bc}\right\} \eta _{bd}e_{a}^{\;\mu }\left(
\Lambda _{\;c}^{d}-\Lambda _{c}^{\;d}\right) \psi ed^{4}x \nonumber \\
&&+\frac{i}{16}\int \delta \left[ f\right] \partial _{\mu }f\overline{\psi }%
\Lambda _{\overline{\psi }}\left\{ \gamma ^{a},\sigma ^{bc}\right\} \Lambda
_{\psi }\eta _{bd}e_{i}^{\;\mu }\Lambda _{a}^{\;i}\left( \Lambda
_{c}^{\;d}-\Lambda _{\;c}^{d}\right) \psi ed^{4}x \nonumber \\
&&-\frac{i}{16}\int \delta \left[ f\right] \partial _{\mu }f\overline{\psi }%
\Lambda _{\overline{\psi }}\left\{ \gamma ^{a},\sigma ^{bc}\right\} \Lambda
_{\psi }\eta _{bd}e_{i}^{\;\mu }\Lambda _{c}^{\;i}\left( \delta
_{a}^{d}+\Lambda _{a}^{\;d}+\Lambda _{\;a}^{d}\right) \psi ed^{4}x \nonumber \\
&&+\frac{i}{16}\int \delta \left[ f\right] \partial _{\mu }f\overline{\psi }%
\Lambda _{\overline{\psi }}\left\{ \gamma ^{a},\sigma ^{bc}\right\} \Lambda
_{\psi }\eta _{bd}\Lambda _{\;i}^{d}e^{i\mu }\left( \eta _{cj}\Lambda
_{a}^{\;j}+\eta _{aj}\Lambda _{c}^{\;j}\right) \psi ed^{4}x, \nonumber
\end{eqnarray}
where $\eta _{ab}$ is the Minkowski metric tensor (+,-,-,-), $\Lambda _{\;a}^{b}
$ are the inverses of $\Lambda _{a}^{\;b} $, and we note
that the expressions $\Theta \left[ \pm f\right] \delta \left[ f\right]
 \partial _{\mu }f$
integrate the same as $\frac{1}{2}\delta \left[ f\right] \partial _{\mu }f$
.

We see that if $\Lambda _{\psi }=\Lambda _{\overline{\psi }}=I$, then $%
\delta S_{D}=0$. \ This means that the introduction of $f$ is not enough to
gauge CPT; we must also include local Lorentz rotations in order to obtain a
nonvanishing $\delta S$ under local CPT transformations. \ The introduction
of local Lorentz rotations requires the introduction of the metric spin
connection $\omega _{\mu ab}$ as noted above.

We now show that the introduction of the local CPT transformations unveils
new physical phenomena distinct, yet coupled, to general relativity. \ If $%
\delta S_{D}=0$ identically, then nothing new is going on other than
defining the CPT symmetry locally on a curved manifold. \ If $\delta
S_{D}\neq 0$ but $\delta S_{D}+\delta S_{\omega }=0$, then the local CPT
transformations are just a part of general relativity without any new
physics. \ If $\delta S_{D}\neq 0$ and $\delta S_{D}+\delta S_{\omega }\neq
0 $, then general relativity cannot accommodate the local CPT symmetry. \ We
then introduce a new gauge field $X_{\mu }$ to arrive at an expanded action
invariant under local $CPT\Lambda $ transformations. \ The proof that there
exists at least one transformation such that $\delta S_{D}\neq 0$ and $%
\delta S_{D}+\delta S_{\omega }\neq 0$ is done by explicit construction
using a simple choice for local Lorentz rotations corresponding to a
velocity boost along the x-axis of Minkowski spacetime.  The nonvanishing
components of $ \Lambda $ for this transformation are $ \Lambda _{2}^{\;2}=
\Lambda _{3}^{\;3}=\Lambda _{\;2}^{2}=\Lambda _{\;3}^{3}=1 $, 
$ \Lambda _{0}^{\;0}=\Lambda _{1}^{\;1}=\Lambda _{\;0}^{0}=\Lambda _{\;1}^{1}
=\cosh \omega $, $ \Lambda _{0}^{\;1}=\Lambda _{1}^{\;0}= -\Lambda _{\;0}^{1}
= -\Lambda _{\;1}^{0}= \sinh \omega $, $\Lambda _{I}=\cosh \frac{\omega }{2}
$, and $\Lambda _{10}= \sinh \frac{\omega }{2} $; where $\Lambda _{I}$,
$\Lambda _{5}$, and $\Lambda _{ab}$ are defined by the
expansion of $\Lambda _{\psi }$: \ $\Lambda _{\psi }=\Lambda _{I}I+\Lambda
_{5}\gamma ^{5}+\Lambda _{ab}\sigma ^{ab}$.  It is
straightforward to show that such a $\Lambda $ applied in regions where $f>0$
satisfies the above criteria that new physical phenomena is unveiled by
gauge $CPT\Lambda $. \ Under this transformation one obtains from eqs. (7)
and (8):%
\[
\delta S_{D}=-\sinh \frac{\omega }{2}\int \delta \left[ f\right] \partial
_{\mu }f\overline{\psi }\left( e_{0}^{\;\mu }\gamma ^{5}\gamma
^{1}+e_{1}^{\;\mu }\gamma ^{5}\gamma ^{0}\right) \psi ed^{4}x, 
\]%
and%
\[
\delta S_{\omega }=\frac{3}{64}\sinh \omega \left( \cosh \omega -1\right)
\int \delta \left[ f\right] \partial _{\mu }f\overline{\psi }\left(
e_{2}^{\;\mu }\gamma ^{5}\gamma ^{3}-e_{3}^{\;\mu }\gamma ^{5}\gamma
^{2}\right) \psi ed^{4}x, 
\]%
where $\omega =\tanh ^{-1}\left( \frac{v}{c}\right) $, $v$ and $c$ being the
velocities of the boost and light respectively. \ Because the $\gamma
^{5}\gamma ^{a}$ are linearly independent, we see that $\delta S_{D}\neq 0$, 
$\delta S_{\omega }\neq 0$, and $\delta S_{D}+\delta S_{\omega }\neq 0$ for
this choice of transformation. \ The fact that $\delta S_{\omega }\neq 0$
means that general relativity is not invariant under local $CPT\Lambda $
transformations. \ Therefore, the new gauge field $X_{\mu }$ must also
compensate for the inhomogeneous (i.e. $\delta \left[ f\right] $) terms
arising from the transformation of $\omega _{\mu ab}$ under local $%
CPT\Lambda $ transformations. \ Thus, we see from variational arguments that
the new gauge field and general relativity are coupled.

One might be concerned about the appearance of the two types of
discontinuities - $\delta \left[ f\right] $ and $\Theta \left[ \pm f\right] $
- within the framework of the calculus of variations. \ Now is an
appropriate point to address this issue because it allows for a summary of
what we have done and where we are going.

We first discuss the appearance of $\delta \left[ f\right] $
discontinuities. \ The only point where $\delta \left[ f\right] $ terms can
be considered as part of the variational theory is at the beginning, where
it is shown that the initial action integrals ($S_{D},S_{\omega
},S_{D}+S_{\omega }$) are not invariant under the local $CPT\Lambda $
transformations. \ Definite integrals are not limited to continuous or
differentiable functions. \ Because the natural setting of delta functionals
is under the definite integral sign, the discontinuities present when
evaluating $\delta S_{D}$, etc. present no mathematical issues in evaluating
the various transformed action integrals. \ Discontinuities containing $%
\Theta \left[ \pm f\right] \delta \left[ f\right] $ also pose no problem
because they integrate the same as $\frac{1}{2}\delta \left[ f\right] $ (see
appendix A). \ Most importantly, the \textit{concept} of testing for a
variation of an action integral under local transformations is not changed
by the \textit{use} of the local $CPT\Lambda $ transformations. \ 

After we introduce the new gauge field $X_{\mu }$, we then construct
functions of the fields which are \textit{required} to be free from any
appearance of $\delta \left[ f\right] $. \ Once these functions of the
fields (e.g., the curvature scalar $R$ formed from $\omega _{\mu ab}$) are
found, \textit{then} the usual machinery of the calculus of variations can
be used without any problems - the $\delta \left[ f\right] $ discontinuities
are not present in the Lagrangian. \ Indeed, the presence of $\delta \left[ f%
\right] $ terms is precisely the reason we will use for rejecting a mass
term for $X_{\mu }$.

We now discuss the appearance of the second type of discontinuities - $%
\Theta \left[ \pm f\right] $ terms. \ These just cause a \textit{partitioning%
} of the original unvaried action integral into unvaried integrals (where $%
f<0$) and transformed integrals (where $f>0$). \ None of these integrals
contains any discontinuities. \ Once a complete Lagrangian is found such
that the transformed action splits into \textit{only} $\Theta \left[ \pm f%
\right] $ regions without any surface integrals arising from $\delta \left[ f%
\right] $, then we can obtain some important results. \ The $\Theta \left[ f%
\right] $ regions will allow us to find how $X_{\mu }$ transforms under the
global $CPT\Lambda $ transformation. \ The $\Theta \left[ -f\right] $
regions will give us the Lagrangian \textit{density} we are looking for. \
This is the Lagrangian density - free of \textit{any} discontinuities and
presence of $f$ - which can subsequently be used in the standard calculus of
variations to find field equations and conserved quantities. \
Unfortunately, the correct $X_{\mu }$ free-field term remains to be found.

\section{INTRODUCTION OF THE NEW GAUGE \\ FIELD}
In order to create an action invariant under local $CPT\Lambda $
transformations, we postulate the existence of a new gauge field $X_{\mu }$
minimally coupled to $\psi $ and $\overline{\psi }$ via a covariant
derivative including the metric spin connection $\omega _{\mu ab}$. \ We now
turn to the task of determining the transformation equations for $X_{\mu }$
under local $CPT\Lambda $ transformations which will lead to the structure
of $X_{\mu }$.

The first step in determining the transformation of $X_{\mu }$ is to notice
that both $\delta S_{D}$ and $\delta S_{\omega }$ are boundary integrals,
i.e. they contain the terms $\delta \left[ f\right] \partial _{\mu }f$ . \
Also, eq. (4) contains the terms $\Theta \left[ -f\right] \delta \left[ f%
\right] \varsigma _{\mu ab}$ and $\Theta \left[ f\right] \delta \left[ f%
\right] \widetilde{\varsigma }_{\mu ab}$ which need to be cancelled out by
the transformation of $X_{\mu }$ under local $CPT\Lambda $ transformations.
\ Hence, the transformation of $X_{\mu }$ under local $CPT\Lambda $
transformations is postulated to be of the form:%
\begin{equation}
X_{\mu }\rightarrow \Theta \left[ -f\right] X_{\mu }+\Theta \left[ f\right] 
\widetilde{X}_{\mu }+\Theta \left[ -f\right] \delta \left[ f\right] Y_{\mu
}+\Theta \left[ f\right] \delta \left[ f\right] \widetilde{Y}_{\mu }.
\end{equation}%
One could also add a term $\delta \left[ f\right] Z_{\mu }$ to the
transformation of $X_{\mu }$. \ It can be shown that such a term does not
eliminate the need for the $\Theta \left[ -f\right] \delta \left[ f\right]
Y_{\mu }$ and $\Theta \left[ f\right] \delta \left[ f\right]
 \widetilde{Y}_{\mu }$ terms and would
only seem to add unnecessary complications. \ Hence, the introduction of $%
\delta \left[ f\right] Z_{\mu }$ will not be pursued further.

We introduce the covariant derivatives, $D_{\mu }\psi $, $D_{\mu }\overline{%
\psi }$:%
\begin{equation}
D_{\mu }\psi =\partial _{\mu }\psi +\frac{1}{2}\omega _{\mu ab}\sigma
^{ab}\psi +\beta X_{\mu }\psi ,
\end{equation}%
and%
\[
D_{\mu }\overline{\psi }=\partial _{\mu }\overline{\psi }-\frac{1}{2}\omega
_{\mu ab}\overline{\psi }\sigma ^{ab}+\beta ^{\ast }\overline{\psi }\gamma
^{0}X_{\mu }^{\dagger }\gamma ^{0}, 
\]%
where $\beta $ is the coupling constant. \ The first order theory [2,3]
unveiled only the $\gamma ^{5}$ components of $X_{\mu }$. \ However, from
eqs. (7) and (8), we see that other components are also needed. \ So, we
treat $X_{\mu }$ as a matrix: \ $X_{\mu }=x_{\mu n}\Gamma ^{n}$, where the $%
x_{\mu n}$ are the dynamical components of $X_{\mu }$; and the $\Gamma ^{n}$
are the 16 linearly independent matrices $I$, $\gamma ^{5}$, $\gamma ^{a}$, $%
\gamma ^{5}\gamma ^{a}$, and $\sigma ^{ab}$.

The replacement of $\partial _{\mu }\psi $ and $\partial _{\mu }\overline{%
\psi }$ in eq. (5) by eqs. (10) results in an expanded action, $S_{D\omega
X} $, which will determine $Y_{\mu }$ and $\widetilde{Y}_{\mu }$ upon
requiring $\delta S_{D\omega X}=0$. \ The form of $\widetilde{X}_{\mu }$,
the transformation of $X_{\mu }$ under global $CPT\Lambda $ transformations,
is determined by requiring the $\Theta \left[ f\right] $ term of the
transformed $S_{D\omega X}$ to change sign, just as in (6). \ 

We determine $Y_{\mu }$ and $\widetilde{Y}_{\mu }$ by requiring the
transformation of the expanded action to have no terms containing $\delta %
\left[ f\right] $. \ By simply substituting the transformation
eqs. (2-4, 9, 10)
into eq. (5) and setting the sums of all terms containing $\Theta %
\left[ -f\right] \delta \left[ f\right] $ and $\Theta \left[ f\right] \delta %
\left[ f\right] $ separately to zero, we can straightforwardly solve for $%
Y_{\mu }$ and $\widetilde{Y}_{\mu }$. \ We obtain:
\begin{equation}
Y_{\mu }=\beta ^{-1}\left[ \partial _{\mu }f\left( I-i\gamma ^{5}\Lambda
_{\psi }\right) -\frac{1}{2}\varsigma _{\mu ab}\sigma ^{ab}\right] ,
\end{equation}%
and%
\[
\widetilde{Y}_{\mu }=\beta ^{-1}\left[ \partial _{\mu }f\left( -I-i\gamma
^{5}\Lambda _{\overline{\psi }}\right) -\frac{1}{2}\widetilde{\varsigma }%
_{\mu ab}\sigma ^{ab}\right] . 
\]%
We note that $\gamma ^{5}\Lambda _{\psi }$ and $\gamma ^{5}\Lambda _{%
\overline{\psi }}$ are linear combinations of $I$, $\gamma ^{5}$, and $%
\sigma ^{ab}$; so we see that only eight of the possible 16 $\Gamma ^{n}$
are needed. \ We assume that $X_{\mu }$ has only these eight $x_{\mu n}$
dynamical components:  $X_{\mu }=x_{\mu I}I+x_{\mu 5}\gamma^{5}+
x_{\mu ab}\sigma ^{ab}$.

We now turn our attention to finding $\widetilde{X}_{\mu }$ by again
substituting the transformation eqs. (1-4) and (9) into the expanded action
obtained by replacing $\partial _{\mu }\psi $ and $\partial _{\mu }\overline{%
\psi }$ by $D_{\mu }\psi $ and $D_{\mu }\overline{\psi }$ in eq. (5). \ As
mentioned above, we require that the $\Theta \left[ f\right] $ terms in the
expanded action change sign and cancel out the corresponding unvaried terms
of the expanded action in the regions where $f>0$. \ We obtain from the $%
\Theta \left[ f\right] $ terms:%
\begin{eqnarray*}
&&e_{b}^{\;\mu }\Lambda _{a}^{\;b}\overline{\psi }\Lambda _{\overline{\psi }%
}\gamma ^{a}\left( \left( \partial _{\mu }\Lambda _{\psi }\right) \psi
+\Lambda _{\psi }\partial _{\mu }\psi +\frac{1}{2}\widetilde{\omega }_{\mu
cd}\sigma ^{cd}\Lambda _{\psi }\psi +\beta \gamma ^{5}\widetilde{X}_{\mu
}\gamma ^{5}\Lambda _{\psi }\psi \right) \\
&&-e_{b}^{\;\mu }\Lambda _{a}^{\;b}\left( \partial _{\mu }\overline{\psi }%
\right) \Lambda _{\overline{\psi }}\gamma ^{a}\Lambda _{\psi }\psi
-e_{b}^{\;\mu }\Lambda _{a}^{\;b}\left( \overline{\psi }\partial _{\mu
}\Lambda _{\overline{\psi }}-\frac{1}{2}\widetilde{\omega }_{\mu cd}%
\overline{\psi }\Lambda _{\overline{\psi }}\sigma ^{cd}\right) \gamma
^{a}\Lambda _{\psi }\psi \\
&&-e_{b}^{\;\mu }\Lambda _{a}^{\;b}\beta ^{\ast }\overline{\psi }\gamma
^{5}\Lambda _{\overline{\psi }}\gamma ^{0}\widetilde{X}_{\mu }^{\dagger
}\gamma ^{0}\gamma ^{5}\gamma ^{a}\Lambda _{\psi }\psi \\
&=&e_{a}^{\;\mu }\overline{\psi }\gamma ^{a}\left( \partial _{\mu }\psi +%
\frac{1}{2}\omega _{\mu cd}\sigma ^{cd}\psi +\beta X_{\mu }\psi \right) \\
&&-e_{a}^{\;\mu }\left( \partial _{\mu }\overline{\psi }-\frac{1}{2}\omega
_{\mu cd}\overline{\psi }\sigma ^{cd}+\beta ^{\ast }\overline{\psi }\gamma
^{0}X_{\mu }^{\dagger }\gamma ^{0}\right) \gamma ^{a}\psi .
\end{eqnarray*}%
The above equation naturally splits into three parts:%
\begin{equation}
e_{b}^{\;\mu }\Lambda _{a}^{\;b}\left( \overline{\psi }\Lambda _{\overline{%
\psi }}\gamma ^{a}\Lambda _{\psi }\partial _{\mu }\psi -\left( \partial
_{\mu }\overline{\psi }\right) \Lambda _{\overline{\psi }}\gamma ^{a}\Lambda
_{\psi }\psi \right) =e_{a}^{\;\mu }\left( \overline{\psi }\gamma
^{a}\partial _{\mu }\psi -\left( \partial _{\mu }\overline{\psi }\right)
\gamma ^{a}\psi \right) ,
\end{equation}%
\[
e_{b}^{\;\mu }\Lambda _{a}^{\;b}\overline{\psi }\Lambda _{\overline{\psi }%
}\gamma ^{a}\left( \left( \partial _{\mu }\Lambda _{\psi }\right) \psi +%
\frac{1}{2}\widetilde{\omega }_{\mu cd}\sigma ^{cd}\Lambda _{\psi }\psi
\right) 
\]%
\[
-e_{b}^{\;\mu }\Lambda _{a}^{\;b}\left( \overline{\psi }\partial _{\mu
}\Lambda _{\overline{\psi }}-\frac{1}{2}\widetilde{\omega }_{\mu cd}%
\overline{\psi }\Lambda _{\overline{\psi }}\sigma ^{cd}\right) \gamma
^{a}\Lambda _{\psi }\psi 
\]%
\[
=\frac{1}{2}e_{a}^{\;\mu }\omega _{\mu cd}\overline{\psi }\left\{ \gamma
^{a},\sigma ^{cd}\right\} \psi , 
\]%
and%
\[
e_{b}^{\;\mu }\Lambda _{a}^{\;b}\overline{\psi }\left( \beta \Lambda _{%
\overline{\psi }}\gamma ^{a}\gamma ^{5}\widetilde{X}_{\mu }\gamma
^{5}\Lambda _{\psi }-\beta ^{\ast }\gamma ^{5}\Lambda _{\overline{\psi }%
}\gamma ^{0}\widetilde{X}_{\mu }^{\dagger }\gamma ^{0}\gamma ^{5}\gamma
^{a}\Lambda _{\psi }\right) \psi 
\]%
\[
=e_{a}^{\;\mu }\overline{\psi }\left( \beta \gamma ^{a}X_{\mu }-\beta ^{\ast
}\gamma ^{0}X_{\mu }^{\dagger }\gamma ^{0}\gamma ^{a}\right) \psi . 
\]%
The first equation is just the identity $e_{b}^{\;\mu }\Lambda
_{a}^{\;b}\Lambda _{\overline{\psi }}\gamma ^{a}\Lambda _{\psi
}=e_{a}^{\;\mu }\gamma ^{a}$. \ The second equation is the minimal coupling
condition of general relativity used to compensate for the introduction of
local Lorentz rotations. \ The last is the equation used to determine the
transformation of $X_{\mu }$ under global $CPT\Lambda $ transformations. \
Use of the above identity in the last equation gives us:%
\[
\gamma ^{5}\left( \beta ^{\ast }\Lambda _{\overline{\psi }}\gamma ^{0}%
\widetilde{X}_{\mu }^{\dagger }\gamma ^{0}\Lambda _{\psi }\gamma ^{a}-\beta
\gamma ^{a}\Lambda _{\overline{\psi }}\widetilde{X}_{\mu }\Lambda _{\psi
}\right) \gamma ^{5}=\beta \gamma ^{a}X_{\mu }-\beta ^{\ast }\gamma
^{0}X_{\mu }^{\dagger }\gamma ^{0}\gamma ^{a}. 
\]%
From this result and the assumption that $X_{\mu }$ is a linear combination
of only $I$, $\gamma ^{5}$, and $\sigma ^{ab}$; we obtain:%
\begin{equation}
\widetilde{X}_{\mu }=\gamma ^{5}\Lambda _{\psi }X_{\mu }\Lambda _{\overline{%
\psi }}\gamma ^{5}=\Lambda _{\psi }X_{\mu }\Lambda _{\overline{\psi }}.
\end{equation}%
We temporarily retain the $\gamma ^{5}$ in eq. (13) to emphasize that the
transformations are $CPT\Lambda $ and not merely $\Lambda $.

\section{ISSUES REGARDING THE FREE-FIELD \\ LAGRANGIAN}
First, we examine the possibility of a mass term $MTr\left( X_{\mu }X^{\mu
\dag }\right) $ - which must be invariant under local $CPT\Lambda $
transformations - in the total Lagrangian density. \ Substitution of eqs.
(9), (11), and (13) into the mass term leads to:%
\begin{equation}
MTr\left( X_{\mu }X^{\mu \dag }\right) \rightarrow MTr\left\{ \Theta \left[
-f\right] X_{\mu }X^{\mu \dag }+\Theta \left[ f\right] \Lambda _{\psi
}X_{\mu }\Lambda _{\overline{\psi }}\Lambda _{\overline{\psi }}^{\dag
}X^{\mu \dag }\Lambda _{\psi }^{\dag }\right\}
\end{equation}%
\[
+MTr\left\{ \Theta \left[ -f\right] \delta \left[ f\right] \left( X_{\mu
}Y^{\mu \dag }+Y_{\mu }X^{\mu \dag }\right) \right\} 
\]%
\[
+MTr\left\{ \Theta \left[ f\right] \delta \left[ f\right] \left( \Lambda
_{\psi }X_{\mu }\Lambda _{\overline{\psi }}\widetilde{Y}^{\mu \dag }+%
\widetilde{Y}_{\mu }\Lambda _{\overline{\psi }}^{\dag }X^{\mu \dag }\Lambda
_{\psi }^{\dag }\right) \right\} 
\]%
\[
+MTr\left\{ \Theta \left[ -f\right] \delta \left[ f\right] \delta \left[ f%
\right] Y_{\mu }Y^{\mu \dag }+\Theta \left[ f\right] \delta \left[ f\right]
\delta \left[ f\right] \widetilde{Y}_{\mu }\widetilde{Y}^{\mu \dag }\right\}
. 
\]%
All terms containing $\delta \left[ f\right] $ must vanish if $X_{\mu }$ is
to have a non-zero mass.

We turn our attention to the terms containing $\delta \left[ f\right] \delta %
\left[ f\right] $ which must vanish simply because the product of the delta
functionals is not defined. \ Substitution of the expressions for $Y_{\mu }$
and $\widetilde{Y}_{\mu }$ yields:%
\begin{eqnarray*}
&&\frac{4}{\beta \beta ^{\ast }}M\Theta \left[ -f\right] \delta \left[ f%
\right] \delta \left[ f\right] \left\{ \partial _{\mu }f\partial ^{\mu }f%
\left[ \left( 1-i\Lambda _{5}\right) ^{2}+\left( \Lambda _{I}\right)
^{2}+2\Lambda ^{0b}\Lambda _{b0}\right. \right. \\
&&\left. +\frac{1}{2}\Lambda ^{ab}\Lambda _{ab}\right] +\frac{1}{2}\partial
^{\mu }f\varsigma _{\mu ab}\Lambda _{cd}\left( \eta ^{0c}\varepsilon
^{abd0}-\eta ^{0a}\varepsilon ^{cdb0}\right) +\frac{1}{2}\varsigma _{\mu
}^{~a0}\varsigma _{\quad a}^{\mu 0} \\
&&\left. +\frac{1}{8}\varsigma _{\mu }^{~ab}\varsigma _{~ab}^{\mu }\right\}
\\
&&+\frac{4}{\beta \beta ^{\ast }}M\Theta \left[ f\right] \delta \left[ f%
\right] \delta \left[ f\right] \left\{ \partial _{\mu }f\partial ^{\mu }f%
\left[ \left( 1+i\Lambda _{5}\right) ^{2}+\left( \Lambda _{I}\right)
^{2}+2\Lambda ^{0b}\Lambda _{b0}\right. \right. \\
&&\left. +\frac{1}{2}\Lambda ^{ab}\Lambda _{ab}\right] -\frac{1}{2}\partial
^{\mu }f\widetilde{\varsigma }_{\mu ab}\Lambda _{cd}\left( \eta
^{0c}\varepsilon ^{abd0}-\eta ^{0a}\varepsilon ^{cdb0}\right) +\frac{1}{2}%
\widetilde{\varsigma }_{\mu }^{~a0}\widetilde{\varsigma }_{\quad a}^{\mu 0}
\\
&&\left. +\frac{1}{8}\widetilde{\varsigma }_{\mu }^{~ab}\widetilde{\varsigma 
}_{~ab}^{\mu }\right\} ,
\end{eqnarray*}%
where $\varepsilon ^{abcd}$ is the Levi-Civita tensor.

The above expression is complicated, so we focus on the vierbein-free (pure
gauge - containing only $f$ and $\Lambda $) terms which must vanish
independently of everything else. \ We obtain for the pure gauge piece:%
\begin{eqnarray*}
&&M\frac{2}{\beta \beta ^{\ast }}\delta \left[ f\right] \delta \left[ f%
\right] \partial _{\mu }f\partial ^{\mu }f\left\{ 2\left[ \left( \Lambda
_{I}\right) ^{2}-\left( \Lambda _{5}\right) ^{2}\right] +4\Lambda
^{0a}\Lambda _{a0}+\Lambda _{ab}\Lambda ^{ab}\right. \\
&&+\frac{1}{4}\Lambda _{a}^{~0}\Lambda _{0}^{~a}+\frac{11}{4}-2\Lambda
_{0}^{~0}+\frac{1}{2}\left( \Lambda _{a}^{~a}+\Lambda _{~a}^{a}\right) \\
&&\left. -\frac{1}{4}\eta _{ab}\Lambda _{~0}^{a}\Lambda _{0}^{~b}+\frac{1}{8}%
\eta ^{ab}\eta _{cd}\Lambda _{~b}^{c}\Lambda _{a}^{~d}\right\} ,
\end{eqnarray*}%
where the $\Lambda _{a}^{~b}$ and $\Lambda _{~b}^{a}$ are the spacetime
representations of Lorentz rotations and their inverses. \ This expression
is still too complicated, so we again resort to the special case of $\Lambda 
$ corresponding to x-axis velocity boosts. \ For this case we obtain:%
\[
M\frac{2}{\beta \beta ^{\ast }}\delta \left[ f\right] \delta \left[ f\right]
\partial _{\mu }f\partial ^{\mu }f\left( 5+\frac{1}{4}\cosh 2\omega +2\cosh
\omega \right) . 
\]%
This does not vanish unless $M=0$, so we do not need to consider anything
else in eq. (14) because it contains either $X_{\mu }$ or the vierbein. \
Therefore, we conclude that $X_{\mu }$ is massless.

We now examine the case that the field strength tensor $F_{\mu \nu }$ is the
same as other gauge theories. \ The arguments for this assumption will be
deferred until section 6. \ We obtain for $F_{\mu \nu }=\left[ D_{\mu
},D_{\nu }\right] $:
\begin{equation}
F_{\mu \nu }=\frac{1}{4}\omega _{\mu ab}\omega _{\nu cd}\left[ \sigma
^{ab},\sigma ^{cd}\right] +\frac{1}{2}\left( \partial _{\mu }\omega _{\nu
ab}-\partial _{\nu }\omega _{\mu ab}\right) \sigma ^{ab}
\end{equation}%
\[
+\frac{\beta }{2}\left( \omega _{\mu ab}\left[ \sigma ^{ab},X_{\nu }\right]
-\omega _{\nu ab}\left[ \sigma ^{ab},X_{\mu }\right] \right) +\beta ^{2}%
\left[ X_{\mu },X_{\nu }\right] +\beta \left( \partial _{\mu }X_{\nu
}-\partial _{\nu }X_{\mu }\right) . 
\]%
The first two terms lead to the Einstein-Hilbert Lagrangian, $\kappa R$, of
general relativity by making use of the identity, $\left[ \sigma
^{ab},\sigma ^{cd}\right] =\eta ^{bc}\sigma ^{ad}+\eta ^{ad}\sigma
^{bc}-\eta ^{bd}\sigma ^{ac}-\eta ^{ac}\sigma ^{bd}$. \ One obtains:%
\begin{eqnarray*}
&&\frac{1}{4}\omega _{\mu ab}\omega _{\nu cd}\left[ \sigma ^{ab},\sigma ^{cd}%
\right] +\frac{1}{2}\left( \partial _{\mu }\omega _{\nu ab}-\partial _{\nu
}\omega _{\mu ab}\right) \sigma ^{ab} \\
&=&\frac{1}{2}\left( \partial _{\mu }\omega _{\nu ab}-\partial _{\nu }\omega
_{\mu ab}+\omega _{\mu ac}\omega _{\nu \;b}^{\;c}-\omega _{\nu ac}\omega
_{\mu \;b}^{\;c}\right) \sigma ^{ab},
\end{eqnarray*}%
which is just $\frac{1}{2}R_{\mu \nu ab}\sigma ^{ab}$, where $R_{\mu \nu ab}$
is the Riemann curvature tensor written in terms of the vierbein embedded in 
$\omega _{\rho jk}$ instead of the more familiar $g_{\mu \nu }$, $\Gamma
_{\rho \nu }^{\mu }$, and their derivatives. \ The familiar form of the
curvature tensor is obtained by contraction with the vierbein, $R_{\mu \nu
\rho \sigma }=R_{\mu \nu }^{\;\;mn}e_{m\rho }e_{n\sigma }$, with $R$
following from further contractions. \ The term $\beta ^{2}\left[ X_{\mu
},X_{\nu }\right] +\beta \left( \partial _{\mu }X_{\nu }-\partial _{\nu
}X_{\mu }\right) $ is the Yang-Mills field strength tensor appearing in
quantum field theory. \ The coupling between the new gauge field and general
relativity comprises the remaining part of eq. (15).

Obtaining the free-field Lagrangian is complicated by the fact that the
free-field term of general relativity comes from a contraction containing $%
F_{\mu \nu }$ whereas the Yang-Mills free-field term is proportional to $%
Tr\left( F_{\mu \nu }^{\dagger }F^{\mu \nu }\right) $. \ If we are to
recover general relativity from eq. (15), then it would appear that we
should perform the required contraction of eq. (15). \ The resulting field
equations will give non-propagating equations for $X_{\mu }$. \ It would
seem that to obtain propagating field equations for $X_{\mu }$, we would
have to use the term $Tr\left( F_{\mu \nu }^{\dagger }F^{\mu \nu }\right) $.
\ Unfortunately, this choice would result in the wrong field equations for
the theory of general relativity. \ If we can show that $R$ is invariant
under $CPT\Lambda $ gauge transformations, then we could "peel-off" the
terms in eq. (15) which contain only the metric spin connection terms
leading to $R$ and consider the rest of eq. (15) for use in a Yang-Mills
Lagrangian.

The starting point of the argument is to calculate the transformation of $R$
in terms of the transformation of $g_{\mu \nu }$, $\partial _{\rho }g_{\mu
\nu }$, etc. because these terms transform more simply than $\omega _{\mu
ab} $ or $\partial _{\nu }\omega _{\mu ab}$ (we are viewing $\omega _{\mu
ab} $ as in the second-order formalism). \ We begin with the relation
between the manifold metric tensor and the vierbein, $g^{\mu \nu }=\eta
^{ab}e_{a}^{\;\mu }e_{b}^{\;\nu }$. \ By substituting eq. (1) into this
expression, we obtain under gauge $CPT\Lambda $ transformations $g^{\mu \nu
}\rightarrow \left( \Theta \left[ -f\right] +\Theta \left[ f\right] \right)
g^{\mu \nu }$. \ Differentiation of this expression gives $\partial _{\rho
}g^{\mu \nu }\rightarrow \left( \Theta \left[ -f\right] +\Theta \left[ f%
\right] \right) \partial _{\rho }g^{\mu \nu }$. \ These two expressions give
us the transformation of the Christoffel symbols, $\Gamma _{\nu \rho }^{\mu
}\rightarrow \left( \Theta \left[ -f\right] +\Theta \left[ f\right] \right)
\Gamma _{\nu \rho }^{\mu }$, and curvature scalar, $R\rightarrow \left(
\Theta \left[ -f\right] +\Theta \left[ f\right] \right) R$. \ Hence, $R$ is
invariant under gauge $CPT\Lambda $ transformations except for a removable
singularity along the boundaries where $f=0$. \ This removable singularity
has no effect on the action integral, so we can indeed "peel-off" the terms
in eq. (15) which contain only the metric spin connection terms contributing
to $R$.

We now examine one possible Yang-Mills Lagrangian obtained from eq. (15).
\ One notes
that $Y^{\mu }$ and $\widetilde{Y}^{\mu }$ contain $\varsigma _{~ab}^{\mu
}\sigma ^{ab}$ and $\widetilde{\varsigma }_{~ab}^{\mu }\sigma ^{ab}$ which
contain vierbein terms. \ To avoid additional complicated calculations due
to the vierbein, we retain the spin connection $\omega _{\mu ab}$ in $D_{\mu
}$ ($D_{\mu }=\partial _{\mu }+\frac{1}{2}\omega _{\mu ab}\sigma ^{ab}+\beta
X_{\mu }$) so that $\varsigma _{~ab}^{\mu }\sigma ^{ab}$ and $\widetilde{%
\varsigma }_{~ab}^{\mu }\sigma ^{ab}$ disappear when $X_{\mu }$ is
introduced. \ This choice does not actually "peel-off" the terms
contributing to $R$ but rather assumes that the contribution of the
resulting Yang-Mills Lagrangian to the total Lagrangian is weaker than $%
\kappa R$. \ One obtains after straightforward substitutions and lengthy
calculations:
\begin{eqnarray}
Tr\left\{ F_{\mu \nu }F^{\mu \nu \dag }\right\} & \rightarrow & 
Tr\left\{ \Theta 
\left[ -f\right] F_{\mu \nu }F^{\mu \nu \dag }+\Theta \left[ f\right] 
\widetilde{F}_{\mu \nu }\widetilde{F}^{\mu \nu \dag }\right.
\\
&&+2\Theta \left[ -f\right] \delta \left[ f\right] \partial _{\mu }f\left(
F^{\mu \nu }\left( \widetilde{D}_{\nu }-D_{\nu }+C_{\nu }\right) ^{\dag
}+h.c.\right) \nonumber \\
&&+2\Theta \left[ f\right] \delta \left[ f\right] \partial _{\mu }f\left( 
\widetilde{F}^{\mu \nu }\left( \widetilde{D}_{\nu }-D_{\nu }+\widetilde{C}%
_{\nu }\right) ^{\dag }+h.c.\right) \nonumber \\
&&+2\delta \left[ f\right] \delta \left[ f\right] \partial _{\mu }f\partial
_{\nu }f\left[ g^{\mu \nu }\left( \widetilde{D}_{\rho }-D_{\rho }\right)
\left( \widetilde{D}^{\rho }-D^{\rho }\right) ^{\dag }\right. \nonumber \\
&&\left. -\left( \widetilde{D}^{\nu }-D^{\nu }\right) \left( \widetilde{D}%
^{\mu }-D^{\mu }\right) ^{\dag }\right] \nonumber \\
&&+2\Theta \left[ -f\right] \delta \left[ f\right] \delta \left[ f\right]
\partial _{\mu }f\partial _{\nu }f\left[ g^{\mu \nu }\left( \left( 
\widetilde{D}_{\rho }-D_{\rho }\right) C^{\rho \dag }+h.c.\right) \right.
\nonumber \\
&&\left. +g^{\mu \nu }C_{\rho }C^{\rho \dag }-\left( \left( \widetilde{D}%
^{\nu }-D^{\nu }\right) C^{\mu \dag }+h.c.\right) -C^{\nu }C^{\mu \dag }%
\right] \nonumber \\
&&+2\Theta \left[ f\right] \delta \left[ f\right] \delta \left[ f\right]
\partial _{\mu }f\partial _{\nu }f\left[ g^{\mu \nu }\left( \left( 
\widetilde{D}_{\rho }-D_{\rho }\right) \widetilde{C}^{\rho \dag
}+h.c.\right) \right. \nonumber \\
&&\left. +g^{\mu \nu }\widetilde{C}_{\rho }\widetilde{C}^{\rho \dag }-\left(
\left( \widetilde{D}^{\nu }-D^{\nu }\right) \widetilde{C}^{\mu \dag
}+h.c.\right) -\widetilde{C}^{\nu }\widetilde{C}^{\mu \dag }\right] \nonumber
 \\
&&+2\Theta \left[ -f\right] \left( F_{\mu \nu }\left( \partial ^{\mu }\left[
\Theta \left[ -f\right] \delta \left[ f\right] \partial ^{\nu }fJ+\Theta %
\left[ f\right] \delta \left[ f\right] \partial ^{\nu }f\widetilde{J}\right]
\right) ^{\dag }+h.c.\right) \nonumber 
\end{eqnarray}
\begin{eqnarray*}
&&+2\Theta \left[ f\right] \left( \widetilde{F}_{\mu \nu }\left( \partial
^{\mu }\left[ \Theta \left[ -f\right] \delta \left[ f\right] \partial ^{\nu
}fJ+\Theta \left[ f\right] \delta \left[ f\right] \partial ^{\nu }f%
\widetilde{J}\right] \right) ^{\dag }+h.c.\right) \\
&&+2\delta \left[ f\right] \partial _{\mu }f\left( \left( \widetilde{D}_{\nu
}-D_{\nu }\right) \left( \partial ^{\mu }\left[ \Theta \left[ -f\right]
\delta \left[ f\right] \partial ^{\nu }fJ+\Theta \left[ f\right] \delta %
\left[ f\right] \partial ^{\nu }f\widetilde{J}\right] \right. \right. \\
&&\left. \left. -\partial ^{\nu }\left[ \Theta \left[ -f\right] \delta \left[
f\right] \partial ^{\mu }fJ+\Theta \left[ f\right] \delta \left[ f\right]
\partial ^{\mu }f\widetilde{J}\right] \right) ^{\dag }+h.c.\right) \\
&&+2\Theta \left[ -f\right] \delta \left[ f\right] \partial _{\mu }f\left(
C_{\nu }\left( \partial ^{\mu }\left[ \Theta \left[ -f\right] \delta \left[ f%
\right] \partial ^{\nu }fJ+\Theta \left[ f\right] \delta \left[ f\right]
\partial ^{\nu }f\widetilde{J}\right] \right. \right. \\
&&\left. \left. -\partial ^{\nu }\left[ \Theta \left[ -f\right] \delta \left[
f\right] \partial ^{\mu }fJ+\Theta \left[ f\right] \delta \left[ f\right]
\partial ^{\mu }f\widetilde{J}\right] \right) ^{\dag }+h.c.\right) \\
&&+2\Theta \left[ f\right] \delta \left[ f\right] \partial _{\mu }f\left( 
\widetilde{C}_{\nu }\left( \partial ^{\mu }\left[ \Theta \left[ -f\right]
\delta \left[ f\right] \partial ^{\nu }fJ+\Theta \left[ f\right] \delta %
\left[ f\right] \partial ^{\nu }f\widetilde{J}\right] \right. \right. \\
&&\left. \left. -\partial ^{\nu }\left[ \Theta \left[ -f\right] \delta \left[
f\right] \partial ^{\mu }fJ+\Theta \left[ f\right] \delta \left[ f\right]
\partial ^{\mu }f\widetilde{J}\right] \right) ^{\dag }+h.c.\right) \\
&&+2\left( \partial _{\mu }\left[ \Theta \left[ -f\right] \delta \left[ f%
\right] \partial _{\nu }fJ\right] \right) \left( \partial ^{\mu }\left[
\Theta \left[ -f\right] \delta \left[ f\right] \partial ^{\nu }fJ+\Theta %
\left[ f\right] \delta \left[ f\right] \partial ^{\nu }f\widetilde{J}\right]
\right. \\
&&\left. -\partial ^{\nu }\left[ \Theta \left[ -f\right] \delta \left[ f%
\right] \partial ^{\mu }fJ+\Theta \left[ f\right] \delta \left[ f\right]
\partial ^{\mu }f\widetilde{J}\right] \right) ^{\dag } \\
&&+2\left( \partial _{\mu }\left[ \Theta \left[ f\right] \delta \left[ f%
\right] \partial _{\nu }f\widetilde{J}\right] \right) \left( \partial ^{\mu }%
\left[ \Theta \left[ -f\right] \delta \left[ f\right] \partial ^{\nu
}fJ+\Theta \left[ f\right] \delta \left[ f\right] \partial ^{\nu }f%
\widetilde{J}\right] \right. \\
&&\left. \left. -\partial ^{\nu }\left[ \Theta \left[ -f\right] \delta \left[
f\right] \partial ^{\mu }fJ+\Theta \left[ f\right] \delta \left[ f\right]
\partial ^{\mu }f\widetilde{J}\right] \right) ^{\dag }\right\} ,
\end{eqnarray*}%
where $\left( \Omega +h.c.\right) $ means $\left( \Omega +\Omega ^{\dag
}\right) $ ( $\Omega $ being a generic expression appearing in eq. (16)),
and we have introduced the following definitions:%
\begin{eqnarray*}
C_{\mu } &=&\left[ D_{\mu }-\partial _{\mu },i\gamma ^{5}\Lambda _{\psi }%
\right] \\
\widetilde{C}_{\mu } &=&\left[ \widetilde{D}_{\mu }-\partial _{\mu },i\gamma
^{5}\Lambda _{\overline{\psi }}\right] \\
J &=&I-i\gamma ^{5}\Lambda _{\psi } \\
\widetilde{J} &=&-I-i\gamma ^{5}\Lambda _{\overline{\psi }}.
\end{eqnarray*}

Before simplifying eq. (16), we must discuss the treatment of terms
containing $\left( \partial ^{\mu }\left[ \Theta \left[ \pm f\right] \delta %
\left[ f\right] \partial ^{\nu }fM\right] \right) $ where $M$ consists of
anything which does not include $\Theta $ or $\delta $. \ Using integration
by parts, it is easily seen that%
\[
\left( \partial ^{\mu }\left[ \Theta \left[ \pm f\right] \delta \left[ f%
\right] \partial ^{\nu }fM\right] \right) N=-\Theta \left[ \pm f\right]
\delta \left[ f\right] \partial ^{\nu }fM\left( \partial ^{\mu }N\right) , 
\]%
where $N$ also can be anything which does not include $\Theta $ or $\delta $%
. \ Unfortunately, $\left( \partial ^{\mu }\left[ \Theta \left[ \pm f\right]
\delta \left[ f\right] \partial ^{\nu }fM\right] \right) $ always appears
with things containing $\Theta $ and $\delta $ in eq. (16), some of which
cannot be dealt with using integration by parts. \ We use the expression [4] 
$\partial ^{\mu }\left( \alpha \Delta \right) =\left( \partial ^{\mu }\alpha
\right) \Delta +\alpha \left( \partial ^{\mu }\Delta \right) $, where $%
\Delta $ is a distribution and $\alpha $ is an "ordinary" function, to see
if it can be extended to an $\alpha $ which also includes $\Theta \left[ \pm
f\right] $. \ Setting $\alpha =\Theta \left[ \pm f\right] M$ and $\Delta
=\delta \left[ f\right] \partial ^{\nu }f$\ , one obtains:%
\newpage
\begin{equation}
\partial ^{\mu }\left[ \Theta \left[ \pm f\right] \delta \left[ f\right]
\partial ^{\nu }fM\right] =\left\{ \left( \pm 1\right) \delta \left[ f\right]
\partial ^{\mu }fM+\Theta \left[ \pm f\right] \partial ^{\mu }M\right\}
\delta \left[ f\right] \partial ^{\nu }f
\end{equation}%
\[
+\Theta \left[ \pm f\right] M\partial ^{\mu }\left( \delta \left[ f\right]
\partial ^{\nu }f\right) . 
\]%
As expected, this expression gives the same result for $\left( \partial
^{\mu }\left[ \Theta \left[ \pm f\right] \delta \left[ f\right] \partial
^{\nu }fM\right] \right) N$ as using integration by parts. \ So, we
optimistically promote eq. (17) to the status of an identity for use in
simplifying eq. (16). \ Use of eq. (17) as an identity eventually leads to:%
\begin{equation}
Tr\left\{ F_{\mu \nu }F^{\mu \nu \dag }\right\} \rightarrow Tr\left\{ \Theta 
\left[ -f\right] F_{\mu \nu }F^{\mu \nu \dag }\right. +\Theta \left[ f\right]
\widetilde{F}_{\mu \nu }\widetilde{F}^{\mu \nu \dag }
\end{equation}%
\begin{eqnarray*}
&&+\delta \left[ f\right] \partial _{\mu }f\left( \left( F^{\mu \nu }\left( 
\widetilde{D}_{\nu }-D_{\nu }+C_{\nu }\right) ^{\dag }+\widetilde{F}^{\mu
\nu }\left( \widetilde{D}_{\nu }-D_{\nu }+\widetilde{C}_{\nu }\right) ^{\dag
}\right) +h.c.\right) \\
&&+2\delta \left[ f\right] \delta \left[ f\right] \partial _{\mu }f\partial
_{\nu }f\left( g^{\mu \nu }\left( \widetilde{D}_{\rho }-D_{\rho }\right)
\left( \widetilde{D}^{\rho }-D^{\rho }\right) ^{\dag }\right. \\
&&\left. -\left( \widetilde{D}^{\nu }-D^{\nu }\right) \left( \widetilde{D}%
^{\mu }-D^{\mu }\right) ^{\dag }\right) \\
&&+\delta \left[ f\right] \delta \left[ f\right] \partial _{\mu }f\partial
_{\nu }f\left[ g^{\mu \nu }\left( \left( \widetilde{D}_{\rho }-D_{\rho
}\right) C^{\rho \dag }+h.c.\right) +\right. g^{\mu \nu }C_{\rho }C^{\rho
\dag } \\
&&-\left( \widetilde{D}^{\nu }-D^{\nu }\right) C^{\mu \dag }-C^{\nu }\left( 
\widetilde{D}^{\mu }-D^{\mu }\right) ^{\dag }-C^{\nu }C^{\mu \dag } \\
&&+g^{\mu \nu }\left( \left( \widetilde{D}_{\rho }-D_{\rho }\right) 
\widetilde{C}^{\rho \dag }+h.c.\right) +g^{\mu \nu }\widetilde{C}_{\rho }%
\widetilde{C}^{\rho \dag }-\widetilde{C}^{\nu }\widetilde{C}^{\mu \dag } \\
&&\left. -\left( \widetilde{D}^{\nu }-D^{\nu }\right) \widetilde{C}^{\mu
\dag }-\widetilde{C}^{\nu }\left( \widetilde{D}^{\mu }-D^{\mu }\right)
^{\dag }\right] \\
&&+\delta \left[ f\right] \partial _{\mu }f\left[ \left( \partial _{\nu
}F^{\mu \nu }\right) J^{\dag }+\left( \partial _{\nu }\widetilde{F}^{\mu \nu
}\right) \widetilde{J}^{\dag }+h.c.\right] \\
&&+\delta \left[ f\right] \delta \left[ f\right] \partial _{\mu }f\partial
_{\nu }f\left[ \left( \partial ^{\mu }\left( J+\widetilde{J}\right) \right)
\left( \widetilde{D}^{\nu }-D^{\nu }\right) ^{\dag }\right. \\
&&\left. -g^{\mu \nu }\left( \partial ^{\rho }\left( J+\widetilde{J}\right)
\right) \left( \widetilde{D}_{\rho }-D_{\rho }\right) ^{\dag }+h.c.\right] \\
&&+\delta \left[ f\right] \delta \left[ f\right] \partial _{\mu }f\partial
_{\nu }f\left[ \left( \partial ^{\mu }J\right) C^{\nu \dag }+\left( \partial
^{\mu }\widetilde{J}\right) \widetilde{C}^{\nu \dag }\right. \\
&&\left. -g^{\mu \nu }\left( \left( \partial ^{\rho }J\right) C_{\rho
}^{\dag }+\left( \partial ^{\rho }\widetilde{J}\right) \widetilde{C}_{\rho
}^{\dag }\right) +h.c.\right] \\
&&+\delta \left[ f\right] \delta \left[ f\right] \partial _{\nu }f\left[
\partial _{\mu }J\left( \partial ^{\nu }f\partial ^{\mu }J-\partial ^{\mu
}f\partial ^{\nu }J\right) ^{\dag }\right. \\
&&\left. \left. +\partial _{\mu }\widetilde{J}\left( \partial ^{\nu
}f\partial ^{\mu }\widetilde{J}-\partial ^{\mu }f\partial ^{\nu }\widetilde{J%
}\right) ^{\dag }\right] \right\} .
\end{eqnarray*}%
We again focus our attention on the pure gauge terms appearing in eq. (18).
\ One obtains:%
\newpage
\begin{equation}
Tr\left\{ \delta \left[ f\right] \delta \left[ f\right] \partial _{\nu }f%
\left[ \partial _{\mu }J\left( \partial ^{\nu }f\partial ^{\mu }J-\partial
^{\mu }f\partial ^{\nu }J\right) ^{\dag }\right. \right.
\end{equation}
\begin{eqnarray*}
&&\left. \left. +\partial _{\mu }\widetilde{J}\left( \partial ^{\nu
}f\partial ^{\mu }\widetilde{J}-\partial ^{\mu }f\partial ^{\nu }\widetilde{J%
}\right) ^{\dag }\right] \right\} \\
&=&Tr\left\{ \delta \left[ f\right] \delta \left[ f\right] \partial _{\nu
}f\partial ^{\nu }f\left[ \left( \partial _{\mu }\Lambda _{\psi }\right)
\left( \partial ^{\mu }\Lambda _{\psi }\right) ^{\dag }+\left( \partial
_{\mu }\Lambda _{\overline{\psi }}\right) \left( \partial ^{\mu }\Lambda _{%
\overline{\psi }}\right) ^{\dag }\right] \right\} \\
&&-Tr\left\{ \delta \left[ f\right] \delta \left[ f\right] \partial _{\nu
}f\partial ^{\mu }f\left[ \left( \partial _{\mu }\Lambda _{\psi }\right)
\left( \partial ^{\nu }\Lambda _{\psi }\right) ^{\dag }+\left( \partial
_{\mu }\Lambda _{\overline{\psi }}\right) \left( \partial ^{\nu }\Lambda _{%
\overline{\psi }}\right) ^{\dag }\right] \right\} \\
&=&8\delta \left[ f\right] \delta \left[ f\right] \partial _{\nu }f\partial
^{\nu }f\left\{ \left( \partial _{\mu }\Lambda _{I}\right) \left( \partial
^{\mu }\Lambda _{I}\right) -\left( \partial _{\mu }\Lambda _{5}\right)
\left( \partial ^{\mu }\Lambda _{5}\right) \right\} \\
&&-8\delta \left[ f\right] \delta \left[ f\right] \partial _{\nu }f\partial
^{\mu }f\left\{ \left( \partial _{\mu }\Lambda _{I}\right) \left( \partial
^{\nu }\Lambda _{I}\right) -\left( \partial _{\mu }\Lambda _{5}\right)
\left( \partial ^{\nu }\Lambda _{5}\right) \right\} \\
&&+2Tr\left\{ \delta \left[ f\right] \delta \left[ f\right] \partial _{\nu
}f\partial ^{\nu }f\left( \partial _{\mu }\Lambda _{ab}\sigma ^{ab}\right)
\left( \partial ^{\mu }\Lambda _{cd}\sigma ^{cd}\right) ^{\dag }\right\} \\
&&-2Tr\left\{ \delta \left[ f\right] \delta \left[ f\right] \partial _{\nu
}f\partial ^{\mu }f\left( \partial _{\mu }\Lambda _{ab}\sigma ^{ab}\right)
\left( \partial ^{\nu }\Lambda _{cd}\sigma ^{cd}\right) ^{\dag }\right\} .
\end{eqnarray*}%
This appears not to vanish; however, we return to our special case of $%
\Lambda $ corresponding to (variable) x-axis boosts just to make sure.
\ One obtains for eq. (19):%
\begin{eqnarray*}
0 &\neq &8\delta \left[ f\right] \delta \left[ f\right] \partial _{\mu
}f\partial _{\nu }f\left\{ g^{\mu \nu }\left[ \left( \partial _{\rho }\cosh 
\frac{\omega }{2}\right) \left( \partial ^{\rho }\cosh \frac{\omega }{2}%
\right) \right. \right. \\
&&\left. +\left( \partial _{\rho }\sinh \frac{\omega }{2}\right) \left(
\partial ^{\rho }\sinh \frac{\omega }{2}\right) \right] \\
&&\left. -\left( \partial ^{\mu }\cosh \frac{\omega }{2}\right) \left(
\partial ^{\nu }\cosh \frac{\omega }{2}\right) -\left( \partial ^{\mu }\sinh 
\frac{\omega }{2}\right) \left( \partial ^{\nu }\sinh \frac{\omega }{2}%
\right) \right\} .
\end{eqnarray*}%
Because the pure gauge terms do not vanish, we know that the transformation
of this type of Yang-Mills Lagrangian $Tr\left\{ F_{\mu \nu }F^{\mu \nu \dag
}\right\} $ is not invariant under gauge $CPT\Lambda $ transformations. \
Therefore, we conclude that this choice of Yang-Mills Lagrangian is not the
free-field
Lagrangian for $X_{\mu }$.

\section{EXPERIMENTAL PREDICTIONS}
Unfortunately, because we do not have a complete theory, assumptions based
on heuristic arguments must also be introduced in order to make physical
predictions. \ We use the context of the galactic dark matter problem as a
possible application of the new force, in part, because of the lack of
direct evidence for dark matter. \ It would seem logical to consider the
alternative possibility of a new force which is responsible for the
unexplained motion of galactic material. \ $CPT$ is worth looking at for the
origin of a new force simply because there are no other experimentally
verified fundamental symmetries to turn to. \ The other logical possibility
of a new force along with new matter will not be considered here; the
following discussion assumes no missing types of matter are involved.

Barring a complicated conspiracy between matter and forces, we see that a
new force must produce a mass independent acceleration, i.e., obey the
equivalence principle. \ The general reasoning behind expecting $X_{\mu }$
to produce such a force is presented in the introduction. \ Additionally,
the $x_{\mu ab}$ components argue for the obeyance of the equivalence
principle by $X_{\mu }$ because they are required - in part - to compensate
for the inhomogeneous terms arising from the transformation of $\omega _{\mu
ab}$. \ Specifically, we assume that the $x_{\mu ab}$ components will
reflect the principle of equivalence via the \textit{direct }alteration of $%
\omega _{\mu ab}$ beyond the $x_{\mu ab}$ contribution to $T_{\mu \nu }$. \
It is this modification (viewing $\omega _{\mu ab}$ as in the Palatini
formalism), $\omega _{\mu ab}\rightarrow \omega _{\mu ab}+x_{\mu ab}$,
rather than gravity sourced by dark matter, which appears as a universal
acceleration by modifying the curvature tensor $R_{\mu \nu ab}$. \ Thus, we
are extending the principle of equivalence by including all spacetime
symmetries. \ Experimental predictions based directly upon $x_{\mu I}$ and $%
x_{\mu 5}$ will not be addressed in this paper.

That regions of HI well beyond the stellar disk follow a flat galactic
rotation curve implies that a new force must also be "long range" given
the dearth of known matter beyond the stellar disk. \ The
fact that $X_{\mu }$ is massless rules out a Yukawa potential for the $%
X_{\mu }$ field. \ However, although $X_{\mu }$ is also a spin-1 field
living in four dimensions, we cannot conclude that the new force follows an
inverse square law without knowing the propagator. \ Without the field
equations, we cannot find the propagator.

We argue for a type of Yang-Mills term as part of the $X_{\mu }$ free-field
Lagrangian, even though that is not enough to guarantee an inverse square
law without additional conditions. \ The contribution to the field equations
from a Yang-Mills type of term also provides a clue as to the identity of
the galactic matter which sources the hypothesized $X_{\mu }$ field.

We expect a Yang-Mills type term to survive for the same reason it is
present in other field theories - parallel transport. \ The key is to look
at the terms $Y_{\mu }$ and $\widetilde{Y}_{\mu }$ which appear in the
transformation of $X_{\mu }$. \ The terms $\frac{1}{2}\varsigma _{\mu
ab}\sigma ^{ab}$ and $\frac{1}{2}\widetilde{\varsigma }_{\mu ab}\sigma ^{ab}$
arise from the transformation of $\omega _{\mu ab}$. \ This is part of the
reason for the necessity of the $x_{\mu ab}$ terms. \ Because $\omega _{\mu
ab}$ is used in the parallel transport of $\psi $, we expect $x_{\mu ab}$ to
also appear in parallel transport in order to cancel out $\frac{1}{2}%
\varsigma _{\mu ab}\sigma ^{ab}$ and $\frac{1}{2}\widetilde{\varsigma }_{\mu
ab}\sigma ^{ab}$. \ However, there are additional transformation terms for $%
x_{\mu ab}$ arising from $i\gamma ^{5}\Lambda _{\psi }$ and $i\gamma
^{5}\Lambda _{\overline{\psi }}$. \ The terms $i\gamma ^{5}\Lambda _{\psi }$
and $i\gamma ^{5}\Lambda _{\overline{\psi }}$ also contain the
transformation terms for $x_{\mu I}$ and $x_{\mu 5}$, for example:

\begin{eqnarray*}
Y_{\mu I} &=&\beta ^{-1}\partial _{\mu }f\left[ 1-\left( \hat{\theta}\cdot 
\hat{\omega}\right) \left( \sin \frac{\theta }{2}\sinh \frac{\omega }{2}%
\right) \right] , \\
Y_{\mu 5} &=&\beta ^{-1}\partial _{\mu }f\left( -i\cos \frac{\theta }{2}%
\cosh \frac{\omega }{2}\right) ,
\end{eqnarray*}%
where $\theta $ and $\omega $ are the parameters for rotations and boosts ($%
\hat{\theta}$ and $\hat{\omega}$ are unit vectors). \ These components of $%
Y_{\mu }$ are interesting because they do not vanish in the absence of local
Lorentz rotations, unlike $\delta S_{D}$. \ This indicates that a free-field
term for at least the $x_{\mu I}$ and $x_{\mu 5}$ components contains
something which does not correspond to parallel transport. \ However, the
appearance of the Lorentz rotations in $Y_{\mu ab}$, $Y_{\mu I}$, and $%
Y_{\mu 5}$ also indicates that parallel transport is necessary in order to
find the free-field term. \ Because Yang-Mills terms arise from parallel
transport around a loop, we expect a Yang-Mills term to be a part of the $%
X_{\mu }$ free-field term. \ Also, the fact that the Yang-Mills Lagrangian
(without $\omega _{\mu ab}$) \textit{is} invariant to first order in $%
\Lambda $ [2,3] lends plausibility to the survival of a Yang-Mills term.

If we assume that a Yang-Mills term (with or without $\omega _{\mu ab}$ in $%
D_{\mu }$) survives in the $X_{\mu }$ free-field Lagrangian, then we see
that chirality plays a prominent role in the $X_{\mu }$ field equations. \
One would find the Yang-Mills contribution to the $x_{\mu I}$ and $x_{\mu 5}$
equations of motion to be (see appendix B):%
\[
i\overline{\psi }e_{a}^{~\mu }\gamma ^{a}\psi =4\beta \left( \partial ^{\mu
}x_{I}^{\nu }-\partial ^{\nu }x_{I}^{\mu }\right) _{;\nu }\text{ and }i%
\overline{\psi }e_{a}^{\;\mu }\gamma ^{a}\gamma ^{5}\psi =4\beta \left(
\partial ^{\mu }x_{5}^{\nu }-\partial ^{\nu }x_{5}^{\mu }\right) _{;\nu }%
\text{ ,} 
\]%
where $;\nu $ denotes covariant differentiation using the Christoffel
symbols. \ By simply adding and subtracting these two equations, we can
redefine $x_{\mu I}$ and $x_{\mu 5}$ in terms of new field variables, $%
x_{\mu L}$ and $x_{\mu R}$, whose source terms are the left- and
right-handed chiral terms obtained from $\psi $. \ Even though $x_{\mu L}$
and $x_{\mu R}$ completely decouple from $\omega _{\mu ab}$ (thereby
affecting gravity solely through their contribution to the energy-momentum
tensor $T_{\mu \nu }$), the special gravitational role of chirality appears
via the Yang-Mills $x^{\mu ab}$ source terms: $i\overline{\psi }\sigma
^{ab}e_{c}^{\;\mu }\gamma ^{c}\psi $.

The hypothesis that neutrinos are the source for the new force is motivated
by three observations. \ First, Dirac neutrinos, $\psi _{\nu }$, have fixed
chirality. \ Thus, if one accepts the "chirality postulate", the copious
amount of neutrinos emitted by stars would be the obvious source for the
force obtained from $X_{\mu }$. \ Second, massless particles emitted from a
finite sized source have the same (monopole) $r^{-2}$ distribution as that
of a spiral galactic dark matter halo. \ Third, neutrinos have negligible
interactions with everything, just like dark matter. \ It is important to
note that the negligible interactions play a different, important role
compared to dark matter. \ If the neutrinos did interact appreciably with
matter, then the absorption by the stellar disk would reduce the effect of
the neutrinos and alter the galactic rotation curves. \ We assume that the
small neutrino mass can be neglected in the following discussion; however,
that small mass, as well as the higher order multipole terms in the neutrino
distribution arising from the stellar distribution in spiral galaxies,
afford the opportunity to detect differences from dark matter predictions.

In order to see why neutrinos play a special role, we examine the source
term $i\overline{\psi }\sigma ^{ab}e_{c}^{\;\mu }\gamma ^{c}\psi $. \ We use
the identity $\sigma ^{ab}\gamma ^{c}=\frac{1}{2}\left\{ \gamma ^{c},\sigma
^{ab}\right\} +\frac{1}{2}\left( \eta ^{bc}\gamma ^{a}-\eta ^{ac}\gamma
^{b}\right) $ to rewrite the source term as:%
\[
i\overline{\psi }\sigma ^{ab}e_{c}^{\;\mu }\gamma ^{c}\psi =\frac{i}{2}%
e_{c}^{\;\mu }\overline{\psi }\left\{ \gamma ^{c},\sigma ^{ab}\right\} \psi +%
\frac{i}{2}e^{b\mu }\overline{\psi }\gamma ^{a}\psi -\frac{i}{2}e^{a\mu }%
\overline{\psi }\gamma ^{b}\psi .
\]%
The terms $\overline{\psi }\gamma ^{a}\psi $ and $\overline{\psi }\gamma
^{b}\psi $ are just vector currents. \ The $\overline{\psi }\left\{ \gamma
^{c},\sigma ^{ab}\right\} \psi $ term produces the spin angular momentum
tensor. \ Now, we can see that the "purely matter-matter channel" ($X_{\mu }$
effects from non-neutrino fermions) due to the $x_{\mu ab}$ terms is highly
suppressed, i.e. not observed, for the exact same reasons that we do not
feel a magnetic force from a tree or get electrocuted when climbing one. \
Ordinarily, the vector currents and the spin angular momentum average to
zero in bulk matter; therefore, $x_{\mu ab}$ averages to zero, also.

However, the neutrinos are a different story. \ An observer outside of a
galaxy will experience neutrinos, $\psi _{\nu }$, passing through him/her
from the galaxy. \ The $\psi _{\nu }$ will be dominated by plane waves
directed away from the galaxy and towards the observer. \ The dominant
propagation direction is fixed; therefore, the vector current source terms
for $x_{\mu ab}$ do not average to zero. \ The fixed propagation direction
combined with the fact that the neutrinos have a fixed chirality means that
the spin angular momentum source terms are also fixed and do not average to
zero. \ Therefore, not only do we see the importance of chirality, but we
also see that the $x_{\mu ab}$ terms do not vanish in this case.

The conditions which dictate whether the force is repulsive or attractive
cannot be addressed without the full field equations. \ So, we will settle
with an experimentally verifiable conjecture that the modification of $%
\omega _{\mu ab}$ due to the chirality of neutrinos results in an additional
acceleration towards the source. \ We note the possibility that
antineutrinos could produce an acceleration away from the source.

At this point we have only conjured up an explanation "homomorphic" to the
dark matter hypothesis. \ The dark matter is replaced by the galactic
neutrinos, and the missing gravitational potential is replaced by $X_{\mu }$%
. \ To see if there is any reality to $X_{\mu }$, we need other experiments.
\ We first turn our attention to the proposed $X_{\mu }$ field produced by
our Sun's neutrinos with the hope of explaining the Pioneer anomaly.

We begin with the application of Newton's second law to a particle's radial
motion, $r\left( t\right) $, from the center of the Sun due to gravity and
the attractive force arising from $X_{\mu }$:%
\begin{equation}
m\frac{d^{2}r}{dt^{2}}=-G_{N}M_{s}m\frac{1}{r^{2}}-km\varphi \left( r\right) 
\frac{1}{r^{2}},
\end{equation}%
where $m$ is the particle's mass, $G_{N}$ is Newton's gravitational
constant, $M_{s}$ is the Sun's mass, $k$ is a constant reflecting the
strength of the new force, and $\varphi \left( r\right) $ is a function of
the neutrinos contained within a sphere of radius $r$. \ The zero mass of $%
X_{\mu }$ is assumed to give the inverse square dependence in the term
 $-km\varphi \left(
r\right) \frac{1}{r^{2}}$, while the extension of the equivalence principal
appears via the presence of $m$ in the same term. \ We assume that the Sun
produces a spherically symmetric distribution of neutrinos.

The term $\varphi \left( r\right) $ requires closer examination. \ Because
of the source term for $x^{\mu jk}$ (see appendix B), $\varphi \left( r\right) 
$ is postulated to be related to $\frac{i}{2}\overline{\psi }_{\nu }\sigma
^{jk}e_{a}^{~\mu }\gamma ^{a}\psi _{\nu }$ which is a simple function (for a
plane wavefunction $\psi _{\nu }$ it is proportional to the number and
energy) of the number and energies of the neutrinos contained within a
sphere of radius $r$, i.e. a function of the neutrino luminosity. \ Given a
constant rate of fusion within the Sun and neglecting any neutrino
interactions with anything, we have $\varphi \left( r\right) =\Psi _{s}\frac{%
r}{c}$ for massless neutrinos, where $\Psi _{s}$ is the "luminosity" of $%
\varphi $ of the Sun. \ We now obtain a slightly more illuminating form of
eq. (20):%
\begin{equation}
\frac{d^{2}r}{dt^{2}}=-G_{N}M_{s}\frac{1}{r^{2}}-k\Psi _{s}\frac{1}{cr}.
\end{equation}%
Unfortunately, eq. (21) cannot explain the Pioneer anomaly. \ The problem is
with the anomalous acceleration term $k\Psi _{s}\frac{1}{cr}$ which has a $%
r^{-1}$ dependence instead of the observed constant (!) acceleration  $%
\approx 8\times 10^{-8}$ cm s-2 [5].
\ However, we shall proceed
to estimate the value of $k\Psi _{s}\frac{1}{cr}$ for comparison as well as
for other possible experiments.

Because of the homomorphism between the gauge $CPT\Lambda $ and dark matter
explanations of the galactic rotation curves, we can make use of the
"observed" dark matter parameters of our galaxy to estimate the value of $%
k\Psi _{s}$. \ For simplicity, our galaxy is modelled as a thin, radially
symmetric disc, neglecting the influence of the central stellar bulge as it
contains only about 15 per cent of the total galactic mass [6]
and, presumably,
neutrino luminosity.
\ We begin with the equation which explicitly
describes the homomorphism between the dark matter gravitational attraction
and the attraction due to the force arising from $x_{\mu ab}$ at the edge $%
R_{e}$ of our galaxy:%
\begin{equation}
\frac{G_{N}M_{D}m}{R_{e}^{2}}=\frac{k\Psi _{g}m}{cR_{e}},
\end{equation}%
where $M_{D}$ is the dark matter mass contained within a sphere of radius $%
R_{e}$, $m$ is the mass of a star at the disc edge, and $\Psi _{g}$ is the
luminosity of $\varphi $ from the entire galactic disc. \ The disc edge is
singled out in order to equate the monopole terms of the two forces. \ We
note, for later use, that because of the $r^{-2}$ mass distribution of the
dark matter we have $\frac{M_{D}}{R_{e}}=\frac{M\left( R\right) }{R}$, where 
$M\left( R\right) $ is the mass of the dark halo contained within a radius
of $R$.

Because neutrino luminosity is not measurable for extrasolar sources, we
need to eliminate $\Psi $ in our experimental predictions. \ We know the
photon luminosity, $L$, and $\Psi $ are related for a given star depending
on the age, mass, metallicity, etc. because both ultimately arise from the
same fusion reactions. \ However, we cannot simply assume that because the
Sun is an average star with regards to spectral class that it also is
average with regards to the neutrino luminosity (actually $\varphi $) of the
galactic stellar population. \ In other words, we cannot assume $\frac{\Psi
_{g}}{\Psi _{s}}\approx \frac{L_{g}}{L_{s}}$, where $L_{g}$ and $L_{s}$ are
the photon luminosities of the galactic disk and the Sun respectively. \
Instead, we write $\frac{\Psi _{g}}{\Psi _{s}}=\alpha \frac{L_{g}}{L_{s}}$,
where the proportionality constant $\alpha $ would be $1$ if the Sun is also
an average star with respect to neutrino luminosity. \ Substitution of this
luminosity relationship into eqs. (21, 22) leads to the anomalous Pioneer
acceleration, $a_{p}$:%
\begin{equation}
a_{p}\left( r\right) =k\Psi _{s}\frac{1}{cr}=\frac{G_{N}L_{s}M_{D}}{\alpha
L_{g}R_{e}r}=\frac{G_{N}L_{s}M_{.5}}{\alpha L_{g}R_{.5}}r^{-1},
\end{equation}%
where $M_{.5}$ and $R_{.5}$ are the half-mass and half-mass radius values of
the dark matter halo. \ We set $\alpha =1$ for simplicity; however, we
include results with $\alpha =.1$ because the emission of $\mathrm{Be_{7}}$,
$\mathrm{B_{8}}$,
and $\mathrm{CNO}$ neutrinos occurs in a narrow range of stellar masses [7]
as well
as to take into account the high photon luminosities of the relatively rare
red giants. \ We use the following values taken from [6]: \ $%
L_{s}=3.85\times 10^{26}$ W, $L_{g}=\left( 2.5\pm 1\right) \times
10^{10}L_{s} $, $M_{D}=2_{-1.8}^{+3}\times 10^{12}M_{s}$, $%
R_{.5}=100_{-80}^{+100}$ kpc, and $M_{s}=1.99\times 10^{30}$ kg. \ Arbitrarily
setting $r=45$ au gives $a_{p}\left( 45 \hspace{.25em} \mathrm{au} \right) 
=2.55\times
10^{-11}$ cm s-2 which is far smaller than the Pioneer anomaly. \ By
setting $\alpha =.1$ and using the appropriate uncertainties of $L_{g}$, $%
M_{D}$, and $R_{.5}$ to maximize $a_{p}$, we can obtain $a_{p}\left(
45 \hspace{.25em} \mathrm{au} \right) =5.3\times 10^{-9}$ cm s-2 which is 
only about 7 per cent of
the Pioneer anomaly.

Although the proposed $X_{\mu }$ field does not explain the Pioneer anomaly,
we can generalize eq. (23) for use in making crude estimates for other
possible experiments. \ We simply introduce a proportionality constant, $%
\eta $, which relates the " $\varphi $ luminosity" of a given source,$\Psi $%
, with $\Psi _{s}$: $\Psi =\eta \Psi _{s}$. \ If the given source produces a
spherically symmetric distribution of neutrinos, then the discussion leading
to eq. (23) obviously generalizes to:

\[
a\left( r\right) =\eta \frac{G_{N}L_{s}M_{.5}}{\alpha L_{g}R_{.5}}r^{-1}%
\text{.} 
\]%
We can produce more accurate estimations by replacing the dark matter values
with the observed galactic rotation velocity, $v_{c}$, at the edge ($R$) of
our galaxy via $M\left( R\right) =\frac{v_{c}^{2}R}{G_{N}}$, where $%
v_{c}=220\pm 20$ km s-1 [6]. \ One obtains:

\begin{equation}
a\left( r\right) =\frac{\eta L_{s}v_{c}^{2}}{\alpha L_{g}}r^{-1},
\end{equation}%
where the acceleration is towards the source for neutrinos. \ Use of eq.
(24) gives the "improved" values for $a_{p}$ of $a_{p}\left( 45 \hspace{.25em}
 \mathrm{au} \right)
=2.8\times 10^{-11}$ cm s-2 and using maximizing uncertainties gives $5.6\times
 10^{-10}$ cm s-2.

Because of the large number of antineutrinos produced, we apply eq. (24) to
nuclear reactors. \ We approximate the antineutrino luminosity of a 1 MW
reactor to be $2\times 10^{17} \hspace{.25em} \overline{\nu }$ s-1 at an
 energy of 6 MeV [8], and the neutrino luminosity of the Sun to be
$.023L_{s}$ [7], which
gives $\eta \approx 2.17\times 10^{-20}$ (assuming that the ratio of the $%
\varphi $ luminosities is the same as that of the neutrino/antineutrino
luminosities). \ Setting $r=10$ m from the reactor core in eq. (24) gives $%
a\left( 10 \hspace{.25em} \mathrm{m} \right) =4.1\times 10^{-19}$ cm s-2 
and using the maximizing
uncertainties gives $8.2\times 10^{-18}$ cm s-2. \ For
a 200 MW reactor the values are $a\left( 10 \hspace{.25em} \mathrm{m} 
\right) =8.2\times
10^{-17}$ cm s-2 ($1.66\times 10^{-15}$ cm s-2,
maximum uncertainties). \ Unfortunately, these results preclude any attempts
to track nuclear powered submarines or to detect shielded, clandestine
fissionable material using $X_{\mu }$.

Accelerators can be used to make large amounts of neutrinos and
antineutrinos, so we apply eq. (24) to the muon neutrino flux produced by
the KEK 12 GeV PS [8]. \ Ignoring the $\nu _{\mu }$
mass and only considering
the peak flux produced at 2 GeV, one obtains $\eta \approx 4.09\times
10^{-13} $. \ The acceleration obtained at the beam dump, $r=300$ m, is $%
a\left( 300 \hspace{.25em} \mathrm{m} \right) =2.6\times 10^{-13}$ cm s-2 
($5.2\times 10^{-12}$ cm s-2, maximum uncertainties). \ However, the
accelerator values are inaccurate because the $\nu _{\mu }$ will not have a
spherically symmetric distribution.

Given the difficulty obtaining a detectable effect in the laboratory, we
turn our attention back to solar neutrinos. \ Conceivably, one could use a
probe orbiting the moon to measure the difference in the probe's
acceleration towards the Sun when the Earth is at different positions in its
orbit during the year. \ A lunar probe is used rather than a probe orbiting
the Earth in order to avoid atmospheric remnants affecting the probe's
motion. \ Measurements of the probe's position during new moon phases allows
the moon to be used as a shield from the stream of solar particles acting on
the probe. \ The maximum difference will obviously occur between the summer
and winter solstices which would give $\Delta a_{probe}=4.3\times
10^{-11}$ cm s-2 ($8.5\times 10^{-10}$ cm s-2, maximum
uncertainties). \ These estimates use the mean Earth-moon distance but a
zero probe-moon distance for simplicity. \ Hopefully, the outstanding
techniques used to determine the anomalous Pioneer accelerations can also be
used for this scenario.

\section{CONCLUSION}
If one accepts the transformations (1) - (4), then all that follows is
straightforward, albeit, tedious. \ So, the conclusion will briefly address
issues regarding a suitable free-field Lagrangian.

The first is the sign given to the $\Theta \left[ f\right] \delta \left[ f%
\right] $ terms when $S_{D}$ is transformed, as discussed before introducing
eq. (6). \ The author has never felt completely comfortable with any
argument [2,3] regarding this issue. \ If one postulates that
\textit{all }%
terms in the transformed $S_{D}$ containing $\Theta \left[ f\right] $ should
receive an additional factor of $-1$, then one obtains:%
\[
\delta S_{D}=\frac{1}{4}\int \delta \left[ f\right] \partial _{\mu
}fe_{a}^{~\mu }\overline{\psi }\gamma ^{5}\left( \left\{ \gamma ^{a},\Lambda
_{\psi }\right\} +\left\{ \gamma ^{a},\Lambda _{\overline{\psi }}\right\}
\right) \psi \left\vert e\right\vert d^{4}x. 
\]%
This is interesting because $\delta S_{D}\neq 0$ even when $\Lambda _{\psi
}=\Lambda _{\overline{\psi }}=I$. \ Unfortunately, the calculations
regarding invariance of free-field terms are more complicated and have not
been pursued very far. \ For example, the transformation of the mass term is
sufficiently complicated that the use of a special case, one parameter
Lorentz rotation does not resolve the issue of whether or not $X_{\mu }$ is
massless.

The origins of additional free-field terms which are not of Yang-Mills form
are of obvious interest. \ While it is conceivable that the addition of the
aforementioned $\delta \left[ f\right] Z_{\mu }$ term to eq. (9) might
either restore invariance to a Yang-Mills Lagrangian or require a new
free-field Lagrangian, the discrete nature of the variations suggests
another path to pursue. \ Although the $CPT$ symmetry transformation is not
continuously connected to the identity, an analogous situation exists when
one looks at the action. \ The parameter $f$ can be continuously deformed
from $f<0$ everywhere (no $CPT\Lambda $ anywhere) to $f>0$ everywhere ($%
CPT\Lambda $ applied everywhere) in an infinite number of ways. \ Different
collections of open sets corresponding to where $f<0$ and $f>0$ are produced
during this process; i.e., different topologies are produced. \ So, we have
the possibility of uncovering additional topological information beyond that
which can be obtained from gauge fields based upon continuous symmetries. \
It would seem that the free-field Lagrangian should reflect this.

The most glaring issue is whether or not a Yang-Mills term is invariant
under local $CPT\Lambda $ transformations. \ There are two basic
possibilities to consider - with or without $\omega _{\mu ab}$ as part of $%
D_{\mu }$. \ This doubles if one includes a $\delta \left[ f\right] Z_{\mu }$
term in the transformation of $X_{\mu }$. \ If one includes the possibility
of the additional factor of $-1$ mentioned above, then the number of
Yang-Mills terms to be checked doubles yet again. \ So, there are eight
possible Yang-Mills terms; only the computationally simplest case was
considered in this paper.

\section*{APPENDIX A}
The use of the transformation eqs. (1-4) and their derivatives leads to
terms containing various products of step functions $\Theta \left[ \pm f%
\right] $ and delta functionals $\delta \left[ f\right] $ when calculating
transformed actions, etc. \ We briefly review how to interpret such terms.

We begin with defining the step function:%
\[
\Theta \left[ f\right] =\left\{ 
\begin{array}{ccc}
0 & \text{if} & f<0 \\ 
b & \text{if} & f=0 \\ 
1 & \text{if} & f>0,%
\end{array}%
\right. \text{ \ \ \ } 
\]%
where $b$ is a finite real constant ($b=0$ in this paper). \ With this
definition of $\Theta \left[ f\right] $ it is obvious that we have

\[
\Theta \left[ -f\right] =\left\{ 
\begin{array}{ccc}
1 & \text{if} & f<0 \\ 
b & \text{if} & f=0 \\ 
0 & \text{if} & f>0.%
\end{array}%
\right. 
\]%
From the expressions for $\Theta \left[ \pm f\right] $ we straightforwardly
obtain our first well-defined products:

\[
\Theta ^{n}\left[ f\right] \equiv \left( \Theta \left[ f\right] \right)
^{n}=\left\{ 
\begin{array}{ccc}
0 & \text{if} & f<0 \\ 
b^{n} & \text{if} & f=0 \\ 
1 & \text{if} & f>0%
\end{array}%
\right. 
\]%
and

\[
\Theta ^{n}\left[ -f\right] \equiv \left( \Theta \left[ -f\right] \right)
^{n}=\left\{ 
\begin{array}{ccc}
1 & \text{if} & f<0 \\ 
b^{n} & \text{if} & f=0 \\ 
0 & \text{if} & f>0,%
\end{array}%
\right. 
\]%
where $n$ is a positive integer. \ We see that $\Theta ^{n}\left[ \pm f%
\right] \approx \Theta \left[ \pm f\right] $, the only difference being a
removable singularity at $f=0$. \ We can now obtain our first mixed product:%
\[
\Theta ^{n}\left[ f\right] \Theta ^{m}\left[ -f\right] =\left\{ 
\begin{array}{ccc}
0 & \text{if} & f\neq 0 \\ 
b^{n+m} & \text{if} & f=0,%
\end{array}%
\right. 
\]%
where $m$ is also a positive integer. \ Indeed, $\Theta ^{n}\left[ f\right]
\Theta ^{m}\left[ -f\right] \approx 0$, the only difference again being a
removable singularity at $f=0$. \ Because of the choice $b=0$, we have the
exact results $\Theta ^{n}\left[ \pm f\right] =\Theta \left[ \pm f\right] $
and $\Theta ^{n}\left[ f\right] \Theta ^{m}\left[ -f\right] =0$ in this
paper.

We proceed to products of the type $\Theta ^{n}\left[ \pm f\right] h$ and $%
\Theta ^{n}\left[ f\right] \Theta ^{m}\left[ -f\right] h$, where $h$ is an
"ordinary" (i.e., not containing any $\Theta \left[ \pm f\right] $ or $%
\delta \left[ f\right] $) and "well-behaved" (i.e., bounded, continuous,
etc.) function comprised of spinors, vierbein, etc. \ When appearing as the
integrand of a definite integral, $\Theta ^{n}\left[ \pm f\right] h$ is
equivalent to $\Theta \left[ \pm f\right] h$ because the removable
singularity at $f=0$ has no effect on the integral. \ Similarly, $\Theta ^{n}%
\left[ f\right] \Theta ^{m}\left[ -f\right] h$ appearing in the integrand is
equivalent to $0$ because the removable singularity at $f=0$ has no effect
on the definite integral. \ Both product interpretations are trivially true
for the choice $b=0$.

We are now ready to consider products containing delta functionals. \ The
starting point is the definition of $\delta \left[ x\right] $:%
\[
\int_{-\infty }^{\infty }\delta \left[ x\right] hdx=\lim_{n\rightarrow
\infty }\int_{-\infty }^{\infty }\delta _{n}\left[ x\right] hdx, 
\]%
where $\delta _{n}\left[ x\right] $ is a sequence of well-behaved functions
representing (i.e., approximating) the delta functional; $\delta _{n}\left[ x%
\right] =\left( \frac{n}{\pi }\right) ^{\frac{1}{2}}e^{-nx^{2}}$ being a
common choice, for example.

The first product considered is $\Theta \left[ x\right] \delta \left[ x%
\right] $:%
\[
\int_{-\infty }^{\infty }\Theta \left[ x\right] \delta \left[ x\right]
hdx=\lim_{n\rightarrow \infty }\int_{-\infty }^{\infty }\delta _{n}\left[ x%
\right] \Theta \left[ x\right] hdx. 
\]%
The function $\delta _{n}\left[ x\right] \Theta \left[ x\right] h$ has a
discontinuity at $x=0$. \ If $b=0$ or $1$, then this is a jump
discontinuity; otherwise, this is a jump discontinuity with an additional
removable singularity. \ Because a removable singularity has no effect on a
definite integral, any choice for $b$ will not affect the discussion
regarding the above integral. \ However, we will restrict the choices of $b$
to be $0$ or $1$ for the rest of the appendix because the additional
removable singularity will cause complications when considering the
differentiation of some expressions containing $\Theta \left[ \pm f\right] $
to be discussed below. \ Before considering an arbitrary $h$, let us look at
the special case $h=1$:%
\[
\int_{-\infty }^{\infty }\Theta \left[ x\right] \delta \left[ x\right]
dx=\lim_{n\rightarrow \infty }\int_{-\infty }^{\infty }\delta _{n}\left[ x%
\right] \Theta \left[ x\right] dx=\lim_{n\rightarrow \infty
}\int_{0}^{\infty }\delta _{n}\left[ x\right] dx=\frac{1}{2}\text{,} 
\]
because $\delta _{n}\left[ x\right] =\delta _{n}\left[ -x\right] $.

We now take an aside to examine $\int_{-\infty }^{\infty }\delta _{n}\left[ x%
\right] hdx$ in more detail. \ As $n\rightarrow \infty $, $\int_{-\infty
}^{\infty }\delta _{n}\left[ x\right] hdx\approx \int_{-\epsilon }^{\epsilon
}$ $\delta _{n}\left[ x\right] hdx,$ where $\epsilon $ becomes an
infinitesimal, because the width of $\delta _{n}\left[ x\right] $ decreases
and $h$ is bounded. \ By the continuity of $\delta _{n}\left[ x\right] h$,
we know that $\int_{-\epsilon }^{\epsilon }$ $\delta _{n}\left[ x\right]
hdx=2\epsilon \left( \delta _{n}\left[ x_{v}\right] h\left( x_{v}\right)
\right) $ for some value of $x_{v}$ such that $-\epsilon \leq $ $x_{v}\leq
\epsilon $ with $x_{v}\rightarrow 0$ as $\epsilon \rightarrow 0$, i.e., as $%
n\rightarrow \infty $. \ So, we see that $\int_{-\infty }^{\infty }\delta
_{n}\left[ x\right] hdx\approx 2\epsilon \left( \delta _{n}\left[ x_{v}%
\right] h\left( x_{v}\right) \right) $ as $n\rightarrow \infty $. \ The
introduction of $\Theta \left[ x\right] $ simply changes the limits of
integration: $\int_{-\infty }^{\infty }\Theta \left[ x\right] \delta _{n}%
\left[ x\right] hdx=\int_{0}^{\infty }\delta _{n}\left[ x\right] hdx$. \
Because $\delta _{n}\left[ x\right] =\delta _{n}\left[ -x\right] $ and $%
x_{v}\rightarrow 0$, we see that $\int_{0}^{\infty }\delta _{n}\left[ x%
\right] hdx\approx \epsilon \left( \delta _{n}\left[ x_{v}\right] h\left(
x_{v}\right) \right) $ as $n\rightarrow \infty $. \ Hence, we see that $%
\Theta \left[ f\right] \delta \left[ f\right] $ integrates the same as $%
\frac{1}{2}\delta \left[ f\right] $. \ Obviously, $\Theta \left[ -f\right]
\delta \left[ f\right] $ also integrates the same as $\frac{1}{2}\delta %
\left[ f\right] $.

Another product appearing in calculations is $\Theta \left[ f\right] \Theta %
\left[ -f\right] \delta \left[ f\right] $. \ We again examine the
corresponding expression:%
\[
\int_{-\infty }^{\infty }\Theta \left[ x\right] \Theta \left[ -x\right]
\delta \left[ x\right] hdx=\lim_{n\rightarrow \infty }\int_{-\infty
}^{\infty }\delta _{n}\left[ x\right] \Theta \left[ x\right] \Theta \left[ -x%
\right] hdx. 
\]%
As discussed above, $\Theta \left[ x\right] \Theta \left[ -x\right]
=0,\forall x\neq 0$ with the possible exception of a removable singularity
at $x=0$. \ Therefore, $\int_{-\infty }^{\infty }\delta _{n}\left[ x\right]
\Theta \left[ x\right] \Theta \left[ -x\right] hdx=0$. \ So, anything in the
calculations which contains $\Theta \left[ f\right] \Theta \left[ -f\right]
\delta \left[ f\right] $ is dropped.

The final product which appears in the calculations is the mathematically
undefined $\delta \left[ f\right] \delta \left[ f\right] $. \ Our
interpretation of any term containing this product is that such a term is
unphysical. \ It is \textit{precisely} because of the appearance of this
product that the $X_{\mu }$ field is massless - the only way to eliminate
this product in the transformation of $MTr\left( X_{\mu }X^{\mu \dag
}\right) $ is to set the mass $M$ equal to $0$. \ The unavoidable appearance
of $\delta \left[ f\right] \delta \left[ f\right] $ in the transformation of
the specific type of $X_{\mu }$ free-field Lagrangian considered in this
paper \textit{is} the reason for rejecting that Lagrangian.

We now examine derivatives of things containing $\Theta \left[ \pm f\right] $
and $\delta \left[ f\right] $ which appear in this paper. \ First, it is
well-known that the derivative of a step function is a delta functional: $\
\partial _{\mu }\left( \Theta \left[ \pm f\right] \right) =\pm \delta \left[
f\right] \partial _{\mu }f$. \ Taking the derivative of a function with a
jump discontinuity is also well-known: \ $\partial _{\mu }\left( \Theta %
\left[ \pm f\right] h\right) =\Theta \left[ \pm f\right] \partial _{\mu
}h\pm h\delta \left[ f\right] \partial _{\mu }f$.

Unless $b=0$, the expression $\partial _{\mu }\left( \Theta \left[ f\right]
\Theta \left[ -f\right] \right) $ is mathematically undefined because that
is trying to take the derivative of a single point. \ This is part of the
reason why $b$ is chosen to be $0$ in this paper.

Finally, we consider $\partial _{\mu }\left( \Theta \left[ \pm f\right]
\delta \left[ f\right] \partial _{\nu }fh\right) $. \ To understand this
term, we look at the expression $\left\{ \partial _{\mu }\left( \Theta \left[
\pm f\right] \delta \left[ f\right] \partial _{\nu }fh\right) \right\} g$
and use integration by parts: \ $\left\{ \partial _{\mu }\left( \Theta \left[
\pm f\right] \delta \left[ f\right] \partial _{\nu }fh\right) \right\}
g=-\Theta \left[ \pm f\right] \delta \left[ f\right] \partial _{\nu
}fh\left( \partial _{\mu }g\right) $. \ This can be treated as above
provided $g$ does not contain $\Theta \left[ \pm f\right] $ or $\delta \left[
f\right] $. \ If $g$ contains $\Theta \left[ \pm f\right] $ or $\delta \left[
f\right] $, then the fatal $\delta \left[ f\right] \delta \left[ f\right] $
terms will arise. \ 

One result which has not been rigorously proven is the use of eq. (17) as an
identity. \ If this equation is not valid, then eq. (16) must be dealt with
again.

\section*{APPENDIX B}
The transformation of the metric spin connection $\omega _{\mu ab}$ under
local $CPT\Lambda $ transformations is given for completeness. \ The
transformation is obtained by simply substituting the transformation of the
vierbein, eq. (1), into the definition of the metric spin connection given
after eq. (4). \ We obtain:%
\begin{eqnarray*}
\widetilde{\omega }_{\mu ab} &=&\frac{1}{2}\left\{ e_{d}^{\;\nu }\left(
\partial _{\mu }e_{c\nu }-\partial _{\nu }e_{c\mu }\right) \left( \Lambda
_{a}^{\;d}\Lambda _{b}^{\;c}-\Lambda _{b}^{\;d}\Lambda _{a}^{\;c}\right)
+\eta _{cd}\left( \Lambda _{a}^{\;d}\partial _{\mu }\Lambda
_{b}^{\;c}-\Lambda _{b}^{\;d}\partial _{\mu }\Lambda _{a}^{\;c}\right)
\right. \\
&&-e_{d}^{\;\nu }e_{c\mu }\left( \Lambda _{a}^{\;d}\partial _{\nu }\Lambda
_{b}^{\;c}-\Lambda _{b}^{\;d}\partial _{\nu }\Lambda _{a}^{\;c}\right)
-\Lambda _{a}^{\;c}\Lambda _{b}^{\;r}\left[ e_{c}^{\;\rho }e_{r}^{\;\sigma
}\left( \partial _{\rho }e_{p\sigma }-\partial _{\sigma }e_{p\rho }\right)
e_{\;\mu }^{p}\right. \\
&&\left. \left. +\left( \partial _{\rho }\Lambda _{p}^{\;s}\right) \Lambda
_{\;d}^{p}\left( \eta _{rs}e_{c}^{\;\rho }-\eta _{cs}e_{r}^{\;\rho }\right)
e_{\;\mu }^{d}\right] \right\} .
\end{eqnarray*}%
\begin{eqnarray*}
\widetilde{\varsigma }_{\mu ab} &=&\frac{1}{2}\left\{ \partial _{\nu }f\left[
e_{d}^{\;\nu }\left( \Lambda _{b}^{\;d}e_{a\mu }-\Lambda _{a}^{\;d}e_{b\mu
}\right) -e_{d}^{\;\nu }e_{c\mu }\left( \Lambda _{a}^{\;d}\Lambda
_{b}^{\;c}-\Lambda _{b}^{\;d}\Lambda _{a}^{\;c}\right) \right. \right. \\
&&\left. -e_{c}^{\;\nu }e_{\;\mu }^{d}\Lambda _{\;d}^{p}\left( \eta
_{pr}\left( \Lambda _{a}^{\;c}\Lambda _{b}^{\;r}-\Lambda _{a}^{\;r}\Lambda
_{b}^{\;c}\right) +\left( \eta _{bp}\Lambda _{a}^{\;c}-\eta _{ap}\Lambda
_{b}^{\;c}\right) \right) \right] \\
&&\left. +\partial _{\mu }f\left( \eta _{bd}\Lambda _{a}^{\;d}-\eta
_{ad}\Lambda _{b}^{\;d}\right) \right\} .
\end{eqnarray*}%
\begin{eqnarray*}
\varsigma _{\mu ab} &=&\frac{1}{2}\left\{ \partial _{\nu }f\left[ 2\left(
e_{a}^{\;\nu }e_{b\mu }-e_{b}^{\;\nu }e_{a\mu }\right) +e_{c\mu }\left(
e_{a}^{\;\nu }\Lambda _{b}^{\;c}-e_{b}^{\;\nu }\Lambda _{a}^{\;c}\right)
\right. \right. \\
&&\left. \left. -e_{\;\mu }^{d}\Lambda _{d}^{\;c}\left( \eta
_{ac}e_{b}^{\;\nu }-\eta _{bc}e_{a}^{\;\nu }\right) \right] -\partial _{\mu
}f\left( \eta _{ac}\Lambda _{b}^{\;c}-\eta _{bc}\Lambda _{a}^{\;c}\right)
\right\} .
\end{eqnarray*}

The equations of motion obtained from a "peel-off" Yang-Mills Lagrangian
density, $H_{\mu \nu }$, are straightforwardly determined by using the
Euler-Lagrange equations on a curved manifold. \ The action used is:%
\begin{eqnarray*}
S &=&\int \{\kappa R-m\overline{\psi }\psi +\frac{i}{2}e_{a}^{\;\mu }%
\overline{\psi }\gamma ^{a}\left( \partial _{\mu }\psi +\frac{1}{2}\omega
_{\mu bc}\sigma ^{bc}\psi +\beta X_{\mu }\psi \right) \}ed^{4}x \\
&&-\int \left\{ \frac{i}{2}e_{a}^{\;\mu }\left( \partial _{\mu }\overline{%
\psi }-\frac{1}{2}\omega _{\mu bc}\overline{\psi }\sigma ^{bc}+\beta ^{\ast }%
\overline{\psi }\gamma ^{0}X_{\mu }^{\dagger }\gamma ^{0}\right) \gamma
^{a}\psi \right\} ed^{4}x \\
&&+\int \left\{ \frac{1}{4}Tr\left( \eta ^{rs}\eta ^{jk}e_{r}^{\;\mu
}e_{s}^{\;\theta }e_{j}^{\;\nu }e_{k}^{\;\phi }H_{\mu \nu }H_{\theta \phi
}^{\dagger }\right) \right\} ed^{4}x,
\end{eqnarray*}%
where%
\[
H_{\mu \nu }=\frac{\beta }{2}\left( \omega _{\mu ab}\left[ \sigma
^{ab},X_{\nu }\right] -\omega _{\nu ab}\left[ \sigma ^{ab},X_{\mu }\right]
\right) +\beta ^{2}\left[ X_{\mu },X_{\nu }\right] +\beta \left( \partial
_{\mu }X_{\nu }-\partial _{\nu }X_{\mu }\right) . 
\]%
We obtain:%
\[
ie_{a}^{\;\mu }\gamma ^{a}\left( \partial _{\mu }\psi +\frac{1}{2}\omega
_{\mu bc}\sigma ^{bc}\psi +\beta X_{\mu }\psi \right) -m\psi =0\text{ (the
Dirac equation),} 
\]%
\begin{eqnarray*}
i\overline{\psi }e_{a}^{~\mu }\gamma ^{a}\psi &=&4\beta \left( \partial
^{\mu }x_{I}^{\nu }-\partial ^{\nu }x_{I}^{\mu }\right) _{;\nu }\text{ and }
\\
i\overline{\psi }e_{a}^{\;\mu }\gamma ^{a}\gamma ^{5}\psi &=&4\beta \left(
\partial ^{\mu }x_{5}^{\nu }-\partial ^{\nu }x_{5}^{\mu }\right) _{;\nu }%
\text{ (the chiral terms of }X_{\mu }\text{),}
\end{eqnarray*}%
\begin{eqnarray*}
i\overline{\psi }\sigma ^{jk}e_{a}^{\;\mu }\gamma ^{a}\psi &=&\left\{ Tr%
\left[ \left( \frac{\beta }{2}\left( \omega ^{\nu ab}\left[ \sigma
_{ab},X^{\mu }\right] -\omega ^{\mu ab}\left[ \sigma _{ab},X^{\nu }\right]
\right) +\beta ^{2}\left[ X^{\nu },X^{\mu }\right] \right. \right. \right. \\
&&\left. \left. \left. +\beta \left( \partial ^{\nu }X^{\mu }-\partial ^{\mu
}X^{\nu }\right) \right) \sigma ^{jk\dag }\right] \right\} _{;\nu } \\
&&-Tr\left[ \left( \frac{\beta }{2}\left( \omega ^{\nu ab}\left[ \sigma
_{ab},X^{\mu }\right] -\omega ^{\mu ab}\left[ \sigma _{ab},X^{\nu }\right]
\right) +\beta ^{2}\left[ X^{\nu },X^{\mu }\right] \right. \right. \\
&&\left. +\beta \left( \partial ^{\nu }X^{\mu }-\partial ^{\mu }X^{\nu
}\right) \right) \left( \frac{1}{2}\omega _{\nu cd}+\beta ^{\ast }x_{\nu
cd}^{\ast }\right) \\
&&\left. \times \gamma ^{0}\left[ \sigma ^{dc},\sigma ^{jk}\right] \gamma
^{0}\right] \text{ (the }x^{\mu jk}\text{ terms of }X^{\mu }\text{),}
\end{eqnarray*}%
\begin{eqnarray*}
\left( e^{q\lambda }e^{p\alpha }-e^{q\alpha }e^{p\lambda }\right) _{;\alpha
} &=&\omega _{\mu }^{\;pn}\left( e_{n}^{\;\lambda }e^{q\mu }-e^{q\lambda
}e_{n}^{\;\mu }\right) +\omega _{\mu }^{\;nq}\left( e_{n}^{\;\lambda
}e^{p\mu }-e_{n}^{\;\mu }e^{p\lambda }\right) \\
&&+\frac{i}{4}\overline{\psi }\left\{ \gamma ^{\lambda },\sigma
^{qp}\right\} \psi \\
&&+\frac{1}{4}Tr\left\{ \left[ \beta \left[ \sigma ^{qp},X^{\mu }\right]
\left( \frac{\beta ^{\ast }}{2}\left( \omega _{\;ab}^{\lambda }\left[ X_{\mu
}^{\dag },\sigma ^{ab\dag }\right] \right. \right. \right. \right. \\
&&\left. -\omega _{\mu ab}\left[ X^{\lambda \dag },\sigma ^{ab\dag }\right]
\right) +\beta ^{\ast 2}\left[ X_{\mu }^{\dag },X^{\lambda \dag }\right] \\
&&\left. \left. \left. +\beta ^{\ast }g^{\lambda \nu }\left( \partial _{\nu
}X_{\mu }^{\dag }-\partial _{\mu }X_{\nu }^{\dag }\right) \right) \right]
+h.c.\right\} \text{ (spin connection)}
\end{eqnarray*}%
and,%
\[
R_{\mu \nu }-\frac{1}{2}Rg_{\mu \nu }=-\kappa ^{2}T_{\mu \nu }\text{
(general relativity).} 
\]

We see that the $x_{\mu I}$ and $x_{\mu 5}$ components of $X_{\mu }$
decouple from the rest of $X_{\mu }$ and $\omega _{\mu ab}$. \ Also, the
equation for $x_{\mu 5}$ is valid only for a massless Dirac field which
suggests a quantum anomaly [9,10] or a missing term from the
Lagrangian.

\section*{ACKNOWLEDGEMENTS}
The author is deeply indebted to numerous friends and family members for
their support and interest. \ However, in the interests of privacy, 
the author will restrict explicit thanks to physicists: \
Dr. Fred Gray.

This paper is dedicated to the memory of Professor Jeeva Anandan.

\section*{REFERENCES}
\begin{enumerate}
\item R. Utiyama, 1956 \textit{Physical Review} \textbf{101} 1597
\item K. Koltko, \textit{Attempts to Find Additional Dynamical Degrees
of Freedom in Spacetime Using Topological and Geometric Methods}, 2000
\textit{Ph.D. Thesis, University of South Carolina}
\item K. Koltko, 2002 \textit{Foundations of Physics Letters} \textbf{15} 299
\item A. Friedman, \textit{Generalized Functions and Partial Differential
Equations}, 2005 \textit{Dover Publications, Inc.} 73
\item J. D. Anderson, P. A. Laing, E. L. Lau, A. S. Liu, M. Nieto, S.
Turyshev; 2002 \textit{Physical Review D} \textbf{65} 082004
\item J. Binney and S. Tremaine, \textit{Galactic Dynamics 2nd. Ed.},
2008 \textit{Princeton University Press} 15, 18, and 771
\item J. N. Bahcall, \textit{Neutrino Astrophysics}, 1989 \textit{Cambridge University
Press} 79 and 166
\item M. Fukugita and T. Yanagida, \textit{Physics of Neutrinos and
Applications to Astrophysics}, 2003 \textit{Springer-Verlag} 143-145 and 147
\item P. O. Mazur, 2000 private communication
\item J. Bekenstein, 2005 private communication
\end{enumerate}

\vspace{1pt}

\end{document}